\documentclass[fleqn,usenatbib]{mnras}
\usepackage{newtxtext,newtxmath}
\usepackage{CJK}
\usepackage[T1]{fontenc}
\DeclareRobustCommand{\VAN}[3]{#2}
\let\VANthebibliography\thebibliography
\def\thebibliography{\DeclareRobustCommand{\VAN}[3]{##3}\VANthebibliography}

\usepackage{graphicx}	
\usepackage{amsmath}	
\usepackage{xcolor}
\usepackage{ulem}

\newcommand{\secref}[1]{Sec.~\ref{sec:#1}}
\newcommand{\beq}{\begin{equation}}
\newcommand{\eeq}{\end{equation}}
\renewcommand{\eqref}[1]{Eq.~(\ref{eq:#1})}

\newcommand{\figref}[1]{Fig.~\ref{fig:#1}}

\renewcommand{\d}{\text{d}}

\title[The impact of baryonic potentials on the gravothermal evolution of SIDM haloes]{The impact of baryonic potentials on the gravothermal evolution of self-interacting dark matter haloes}

\author[Y.-M. Zhong, D. Yang \& H.-B. Yu]{
Yi-Ming Zhong,$^{1,2}$\thanks{E-mail: yimingzhong@uchicago.edu}
Daneng Yang,$^{3}$\thanks{E-mail: daneng.yang@ucr.edu} and
Hai-Bo Yu$^{3}$\thanks{E-mail: haiboyu@ucr.edu}
\\
$^{1}$Kavli Institute for Cosmological Physics, University of Chicago, Chicago, IL 60637, USA\\
$^{2}$Department of Physics, City University of Hong Kong, Tat Chee Avenue, Kowloon, Hong Kong SAR, China\\
$^{3}$Department of Physics and Astronomy, University of California, Riverside, California 92521, USA
}

\pubyear{2023}

\begin{document}
\label{firstpage}
\pagerange{\pageref{firstpage}--\pageref{lastpage}}
\maketitle

\begin{abstract}

The presence of a central baryonic potential can have a significant impact on the gravothermal evolution of self-interacting dark matter (SIDM) haloes. We extend a semi-analytical fluid model to incorporate the influence of a static baryonic potential and calibrate it using controlled N-body simulations. We construct benchmark scenarios with varying baryon concentrations and different SIDM models, including constant and velocity-dependent self-interacting cross sections. The presence of the baryonic potential induces changes in SIDM halo properties, including central density, core size, and velocity dispersion, and it accelerates the halo's evolution in both expansion and collapse phases. Furthermore, we observe a quasi-universality in the gravothermal evolution of SIDM haloes with the baryonic potential, resembling a previously known feature in the absence of the baryons. By appropriately rescaling the physical quantities that characterize the SIDM haloes, the evolution of all our benchmark cases exhibits remarkable similarity. Our findings offer a framework for testing SIDM predictions using observations of galactic systems where baryons play a significant dynamical role.

\end{abstract}


\section{Introduction}

In recent years, there has been growing interest in self-interacting dark matter (SIDM); see~\cite{2018PhR...730....1T} and~\cite{2022arXiv220710638A} for reviews. In this scenario, dark matter particles in a halo can scatter and collide through a new force aside from gravity. Dark matter self-interactions allow efficient energy exchanges between cold inner and hot outer regions of the halo and change the inner density profile accordingly~\citep[e.g.,][]{Spergel:1999mh,Dave:2000ar,Vogelsberger:2012ku,Rocha:2012jg,Vogelsberger:2015gpr,Kaplinghat:2015aga,Robertson:2016qef,Nadler:2020ulu,Fischer:2022rko,Rahimi:2022ymu}. SIDM predicts more diverse dark matter distributions in galaxies, compared to cold, collisionless dark matter (CDM), in better agreement with observations~\citep{Kaplinghat:2015aga,Creasey:2017qxc,Kamada:2016euw,Ren:2018jpt,Kaplinghat:2019svz,Zavala:2019sjk,Santos-Santos:2019vrw,Yang:2020iya,Correa:2020qam,Zeng:2021ldo,Zentner:2022xux,Correa:2022dey,Yang:2022mxl,Nadler:2023nrd}. From the perspective of particle physics, SIDM indicates the existence of a new force mediator, which could be searched in various terrestrial dark matter experiments~\citep{2018PhR...730....1T}. 

The gravothermal evolution of an SIDM halo has two distinct phases. At the first one, the self-interactions transport heat inward, a shallow density core forms, and the core size increases with time. After the halo reaches its maximal core expansion, the heat direction is reversed, and the central density increases, resulting in core collapse~\citep{Balberg:2002ue,Koda:2011yb,Essig:2018pzq,Huo:2019yhk,Nishikawa:2019lsc,Sameie:2019zfo,Kahlhoefer:2019oyt,Turner:2020vlf,Zeng:2021ldo,Outmezguine:2022bhq,Yang:2022hkm,Yang:2022zkd,Yang:2023jwn,Nadler:2023nrd}. The SIDM thermalization provides a mechanism that ties dark matter and baryon distributions in both phases. For example, in the expansion phase, the core size is correlated with the baryon concentration~\citep{Kaplinghat:2013xca,Vogelsberger:2014pda,Creasey:2017qxc, Sameie:2018chj,Robertson:2017mgj,Despali:2018zpw,Jiang:2022aqw}. Compared to the density core induced by baryonic feedback in CDM, the SIDM core is more resilient to star formation history~\citep{Robles:2019mfq,Sameie:2021ang,Vargya:2021qza,Burger:2022cjo}. Furthermore, the presence of the baryons can accelerate the onset of core collapse and shorten the collapse timescale~\citep{Elbert:2016dbb,Sameie:2018chj,Feng:2020kxv,Yang:2023stn}. 

In this work, we investigate the impact of a central baryonic potential on the gravothermal evolution of SIDM haloes. We will extend the conducting fluid model, initially developed for the SIDM-only case~\citep{Balberg:2002ue}, to incorporate effects of a baryonic potential as in~\cite{Feng:2020kxv} and further calibrate it with controlled N-body simulations. We construct benchmarks for initial conditions with varying baryon concentrations, as well as different SIDM models, including constant and velocity-dependent self-interacting cross sections. Our N-body and fluid simulations cover the entire range of SIDM halo evolution. In particular, the calibrated fluid model is well suited for simulating core-collapse haloes as it has high resolution and is flexible.  

We will show that for a given baryonic potential the final SIDM distributions are insensitive to growth history of the potential. Our N-body and fluid simulations reveal detailed evolution trajectories of the central dark matter density, velocity dispersion, and core size for the benchmarks. The presence of the potential can accelerate both core-forming and -collapsing processes of the SIDM haloes, and we derive an analytical formula for estimating the collapse timescale. 

Furthermore, we will demonstrate that the evolution of SIDM haloes exhibits a universal behavior in the presence of the baryonic potential. For a fixed baryonic potential, the explicit dependence on the cross section can be absorbed by rescaling the evolution time with the collapse time. For different baryon concentrations, we introduce a new set of fiducial quantities, and the evolution of the rescaled central density, velocity dispersion, and core size, becomes almost identical for our benchmarks. Our findings can be used for testing SIDM predictions using observations of galactic systems where baryons are dynamically important.

The rest of the paper is organized as follows: In~\secref{simulation}, we discuss details about our N-body and fluid simulations, as well as benchmark cases. In~\secref{results}, we present the simulation results and discuss the influence of the baryonic potential on halo evolution. We discuss the universal behavior of  SIDM halo evolution in~\secref{scaling} and conclude in~\secref{conclusion}. In App.~\ref{sec:convergence}, we discuss convergence tests of our N-body simulations. In App.~\ref{sec:fluidcode}, we show the numerical procedure for performing the fluid simulation. In App.~\ref{sec:hydro}, we derive an analytical solution to the hydrostatic equation. In App.~\ref{sec:linear_relation}, we show the relation between collapse timescale and SIDM cross section found in our fluid simulations. In App.~\ref{sec:nbodyextract}, we provide details about applying Gaussian process regression to obtain the halo properties from our N-body simulations.

\begin{table*}
	\centering
	\caption{Parameters for the benchmark scenarios we simulate (from left to right): labeling name, halo scale density and radius, total baryonic mass, baryonic scale density and radius, the effective cross section, and total central gravitational potential at $t=0$. The labels "baryonM, D,'' and ``C" denote median, diffuse, and compact baryon distributions, respectively. For the constant SIDM models, $\sigma_m=\sigma^{\rm eff}_m$.
	}
	\label{tab:benchmark}
	\begin{tabular}{l|ccccc|c|c}
		\hline
		Name   & $\rho_{s}$ [${\rm M}_\odot/\text{kpc}^3$] &  
		 $r_{s}$ [kpc]& $M_{b,\text{tot}}$ [${\rm M}_\odot$] & $\rho_h$ [${\rm M}_\odot/\text{kpc}^3$] & 
		 $r_h$ [kpc]& $\sigma_m^{\text{eff}}$ [cm$^2$/g] & $\Phi_c$ [$\text{km}^2/\text{s}^2$] \\
		\hline
		SIDM10-only & $6.9\times 10^6$ & $9.1$ & 0 & -- & -- & 10 & $-3.1\times 10^{4} $\\
		SIDM10+baryonM  & $6.9\times 10^6$ & $9.1$ & $1.0\times10^9$ & $3.6\times 10^8$ & 0.77 & 10 & $-3.6\times 10^{4}$\\
		SIDM10+baryonD  & $6.9\times 10^6$ & $9.1$ & $1.0\times10^9$ & $9.0\times 10^7$ & 1.2 & 10 & $-3.4\times 10^{4}$ \\
		SIDM10+baryonC  & $6.9\times 10^6$ & $9.1$ & $2.0\times10^9$ & $5.3\times 10^8$ & 0.85 & 10 & $-4.1\times 10^{4}$\\
				\hline
		SIDM100+baryonM  & $6.9\times 10^6$ & $9.1$ & $1.0\times10^9$ & $3.6\times 10^8$ & 0.77 & 100 & $-3.6\times 10^{4}$ \\
		\hline
		vdSIDM-only  & $6.9\times 10^6$ & $9.1$ & 0 & -- & -- & 9.7 & $-3.1\times 10^{4}$ \\
		vdSIDM+baryonM  & $6.9\times 10^6$ & $9.1$ & $1.0\times10^9$ & $3.6\times 10^8$ & 0.77 & 8.6 & $-3.6\times 10^{4}$\\
		\hline
	\end{tabular}
\end{table*}

\section{Simulations}
\label{sec:simulation}

We use both N-body and fluid simulations to study the impact of a baryonic potential on the evolution of SIDM haloes. The fluid model offers high spatial resolution and is computationally inexpensive, making it suitable for studying haloes in the collapse phase. However, it requires calibration using N-body simulations. We construct benchmark scenarios for initial conditions and evolve them using both simulations. We will further use the calibrated fluid model to study the properties of the central halo that is deeply collapsed.

\subsection{Initial conditions and benchmarks}

We assume the initial halo follows a Navarro-Frenk-White (NFW) density profile~\citep{Navarro:1996gj},
\beq
\rho_\chi (r) = \rho_{s} \left(\frac{r}{r_{s}}\right)^{-1}\left(1+\frac{r}{r_{s}}\right)^{-2},
\label{eq:rho_chi}
\eeq
where $\rho_{s}$ and $r_{s}$ are the scale density and radius, respectively. For the NFW profile, the enclosed mass within $r$ is 
\beq
M_\chi(r) = 4\pi \rho_{s} r_{s}^3 \left[\ln\left(1+\frac{r}{r_{s}}\right) - \frac{r}{r+r_{s}}\right]
\eeq
and the gravitational potential
\beq
\label{eq:nfwpotential}
\Phi_{\chi} (r)  = -\frac{4\pi G \rho_{s} r_{s}^3}{r} \ln\left(1+\frac{r}{r_{s}}\right).
\eeq

For the central baryonic component, we assume a Hernquist density profile~\citep{1990ApJ...356..359H}
\beq
\rho_b (r) =\rho_h \left(\frac{r}{r_h}\right)^{-1} \left(1+ \frac{r}{r_h}\right)^{-3},\quad \rho_h\equiv \frac{M_{b,\text{tot}}}{2\pi r_h^3} 
\eeq
where $M_{b,\text{tot}}$ and $r_h$ are the total mass and the scale radius of the baryonic component, respectively. The enclosed mass is 
\beq
M_b (r) = M_{b,\text{tot}} \left(1 + \frac{r_h}{r}\right)^{-2},
\eeq
and the gravitational potential 
\beq
\Phi_b(r) = - \frac{G M_{b,\text{tot}}}{r+r_h}.
\label{eq:b_pot}
\eeq

The parameters of the halo and baryonic components are summarized in Table~\ref{tab:benchmark}. The halo mass is $M_{200}=1.2\times 10^{11}~\text{M}_{\odot}$, with a median concentration of $c_{200}\approx15$~\citep{Dutton:2014xda}. We take two values for the baryonic mass $M_{b,\text{tot}}=1.0\times10^9~\text{M}_\odot$ and $2.0\times10^{9}~{\text{M}_\odot}$, following the stellar-to-halo mass relation with its median and extreme scatter ($+6\sigma$)~\citep{2013MNRAS.428.3121M}, and three values for the scale radius $r_h=0.77~{\rm kpc}$, $1.2~{\rm kpc}$, and $0.85~{\rm kpc}$ motivated by the stellar size-mass relation~\citep{2019MNRAS.485..382C}. We consider three combinations for the baryonic potential, dubbed as baryonM, D, and C, corresponding to the median, diffuse, and compact baryon distributions, respectively; see Table~\ref{tab:benchmark} for details.

For SIDM, we first focus on constant cross sections $\sigma_m\equiv\sigma/m = 10~\rm cm^2/g$ and $100~\rm cm^2/g$. It is important to note that for a realistic SIDM model, the cross section must be velocity-dependent, and its value diminishes to $\sim0.1~{\rm cm^2/g}$ towards scales of galaxy cluster~\citep{Kaplinghat:2015aga,Andrade:2020lqq,Sagunski:2020spe}. Thus, the constant $\sigma_m$ value we take should be regarded as an effective cross section for the halo~\citep{Yang:2022hkm}. In addition, we consider an SIDM model with Rutherford scattering~\citep{feng:2009mn,Ibe:2009mk,Tulin:2013teo}, which is velocity- and angular-dependent 
\beq
\frac{\d\sigma}{\d \cos\theta} = \frac{\sigma_{0}w^4}{2[w^2+{v_\text{rel}^{2}}\sin^2(\theta/2)]^2 },
\label{eq:diff}
\eeq
where $\theta$ and $v_\text{rel}$ are the scattering angle and the relative velocity of incoming dark matter particles in the centre of the mass frame, respectively. We choose the parameters $\sigma_0/m=100~\rm cm^2/g$ and $w=75.3~\rm km/s$ such that the corresponding effective constant cross section is $\sim10~{\rm cm^2/g}$ for the halo we consider in this work~\citep[e.g.,][Eq.~(4.2)]{Yang:2022hkm}. For velocity-dependent SIDM (vdSIDM), we only consider the medium baryon distribution (baryonM), as well as the SIDM-only case; see Table~\ref{tab:benchmark}.

\subsection{N-body simulations}
\label{sec:nbody}

For the N-body simulations, we use the public~\texttt{GADGET-2} program~\citep{Springel:2005mi} with an SIDM module developed in~\cite{Yang:2022hkm}. The module uses similar numerical techniques as in~\cite{Robertson:2016qef,Robertson:2016xjh}. It can simulate both constant and vdSIDM models, including velocity- and angular-dependence in~\eqref{diff}. The total number of simulation particles is $4\times10^6$, and their mass is $3\times10^4~{\rm M_\odot}$. The force softening length is $h=0.13~{\rm kpc}$. 

\begin{figure}
	\includegraphics[width=\columnwidth]{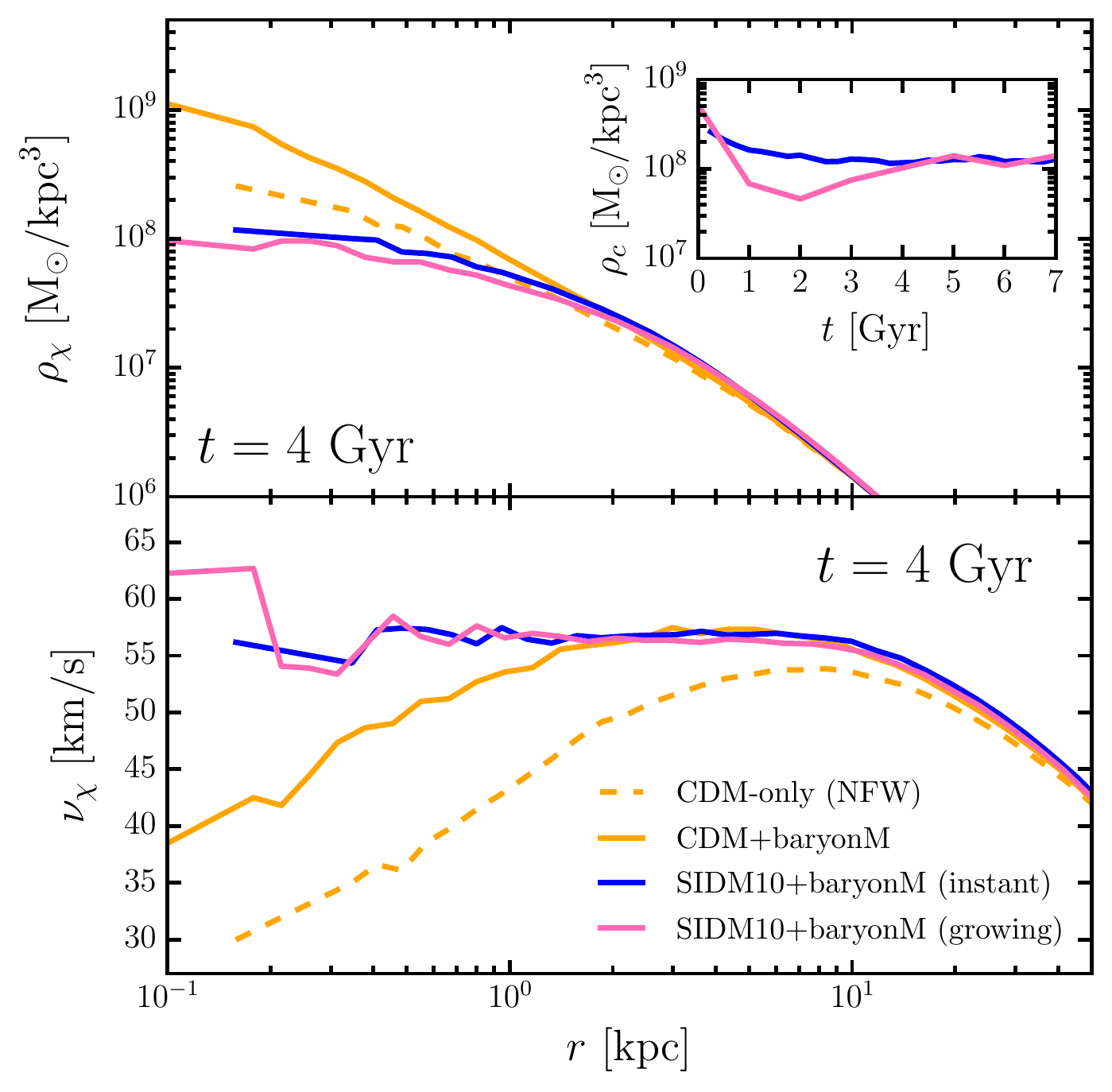}
    \caption{{\bf Top:} Dark matter density profiles at $t=4~{\rm Gyr}$ for SIDM10+baryonM with instant (blue) and growing (magenta) baryonic potentials. The inset displays the evolution of the dark matter central density $\rho_c$ for both approaches. {\bf Bottom:}  Corresponding 1D dark matter velocity-dispersion profiles. For comparison, CDM simulations with the instantaneous potential (solid orange), as well as the initial NFW halo (dashed orange), are also shown. 
    }
    \label{fig:grow} 
\end{figure}

We have tested two approaches to implementing the baryonic potential with N-body simulations: inserting the potential instantaneously at $t=0$; growing it linearly in mass from $t=0$ to $t=4~{\rm Gyr}$. Fig.~\ref{fig:grow} shows dark matter density (top) and velocity-dispersion (bottom) profiles evaluated at $t=4~{\rm Gyr}$ for SIDM10+baryonM with instant (blue) and growing (magenta) baryonic potentials. The insert in the top panel denotes the evolution of the central density. We see that the difference in $\rho_\chi$ and $\nu_\chi$ from the two approaches becomes negligible for $t\gtrsim4~{\rm Gyr}$, as dark matter self-interactions thermalize the inner halo quickly. For comparison, we also show the results for CDM with the instant insertion (solid orange), as well as the initial NFW profile (dashed orange). As expected, the central density of the CDM halo is enhanced due to baryonic concentration~\citep{Blumenthal:1985qy}.

Our test demonstrates that for a given baryonic potential the final SIDM distributions are insensitive to growth history of the potential, because the self-interactions lead to rapid thermalization of the inner halo. For the rest of this work, we will take the instant approach in both N-body and fluid simulations, as we discuss next. 

\subsection{Fluid simulations}
\label{sec:fluid}
For the fluid simulations, we use a set of differential equations that describe a hydrostatic equilibrium system 
\begin{align}
\frac{\partial M_{\chi}}{\partial r} ={}& 4\pi r^2 \rho_\chi,\nonumber\\
\frac{\partial (\rho_\chi \nu_\chi^2)}{\partial r} ={}&- \rho_\chi \frac{\partial (\Phi_\chi+\Phi_b)}{\partial r} = - \frac{G (M_\chi+M_{b}) \rho_\chi}{r^2},\nonumber\\
\frac{\partial L_\chi}{\partial r} ={}& -4 \pi \rho_\chi r^2 \nu_\chi^2 D_t \ln \frac{\nu_\chi^3}{\rho_\chi},\nonumber\\
\frac{L_\chi}{4\pi r^2}={}&-\kappa\frac{\partial (m\nu^2_\chi/k)}{\partial r},
\label{eq:fluid}
\end{align}
where $L_\chi$ is the luminosity profile, $k$ is the Boltzmann constant and $D_t$ denotes the Lagrangian time derivative.~\cite{Balberg:2002ue} first introduced this concluding fluid model to study the evolution of an SIDM halo, and~\cite{Feng:2020kxv} extended it to include a baryonic potential. Heat conductivity of the SIDM fluid, $\kappa$, can be expressed as $\kappa=(\kappa^{-1}_{\rm lmfp}+\kappa^{-1}_{\rm smfp})^{-1}$~\citep{Balberg:2002ue},
where $\kappa_{\rm lmfp}\approx0.27 \beta \rho_\chi\nu^3_\chi\sigma_m k/(Gm)$ and $\kappa_{\rm smfp}\approx2.1\nu_\chi k/(m \sigma_m)$ are the conductivity in the long- and short-mean-free-path regimes, respectively. We determine the conduction coefficient $\beta$, an ${\cal O}(1)$ factor, using calibration against N-body simulations. For the SIDM-only case, $\beta\simeq 0.6\textup{--} 0.75$~\citep{Koda:2011yb,Pollack:2014rja, Essig:2018pzq,Nishikawa:2019lsc}. We recalibrate it with our controlled N-body simulations including the baryonic potential. In addition, the boundary conditions are $M_\chi = 0$ at $r=0$, $M_\chi = M_{\chi,\text{tot}}$ and $L_\chi =0$ at the halo boundary. 

We follow the numerical recipe as in~\cite{Balberg:2002ue,Feng:2020kxv}. For each of the physical quantities in~\eqref{fluid}, we divide it by its fiducial value, see~Table~\ref{tab:fiducial}, and then convert the set of equations in~\eqref{fluid} into the dimensionless form 
\begin{align}
\frac{\partial \hat M_{\chi}}{\partial \hat r} ={}&\hat r^2 \hat \rho_\chi,
~
\frac{\partial (\hat \rho_\chi \hat \nu_\chi^2)}{\partial \hat r} = -\frac{(\hat M_\chi + \hat M_b)\hat \rho_\chi}{\hat r^2}, \nonumber\\
\frac{\partial \hat L_\chi}{\partial \hat r} ={}& - \hat \rho_\chi \hat r^2 \hat \nu_\chi^2 D_{\hat t} \ln \frac{\hat \nu^3_\chi}{\hat \rho_\chi}, \nonumber\\
\frac{\hat L_\chi}{\hat r^2} ={}& - \left[(3.4 \beta \hat \rho_\chi \hat \nu^3 \hat \sigma_m)^{-1}+ \left(\frac{2.1 \hat \nu}{\hat \sigma_m}\right)^{-1}\right]^{-1}\frac{\partial \hat \nu_\chi^2}{\partial \hat r}.
\label{eq:fluiddimless}
\end{align}
We segregate the halo into a series of radial Lagrangian zones and iterate ``conduction-then-relaxation'' steps to model SIDM halo evolution; see App.~\ref{sec:fluidcode} for details.

\begin{table}
	\centering
	\caption{Fiducial quantities used in the fluid simulations.
	}
	\label{tab:fiducial}
	\begin{tabular}{ccc} 
		\hline
	Fiducial quantity & Expression & Value \\
	\hline
	$\rho_0$ & $\rho_{s}$ & $6.9 \times 10^6\,\text{M}_\odot/\text{kpc}^3$ \\
	$r_0$ & $r_{s}$ & $9.1\,\text{kpc}$\\
    $M_0$  & $4\pi \rho_{s} r_{s}^3$ & $6.5\times 10^{10}\,\text{M}_\odot$\\
    $\sigma_{m,0}$ & $(r_{s} \rho_{s})^{-1}$ & $76\,\text{cm}^2/\text{g}$\\
    $\nu_0$ & $(4\pi G \rho_{s})^{1/2} r_{s}$ & $176\,\text{km}/\text{s}$\\
    $L_0$  & $(4\pi)^{5/2} G^{3/2} \rho_{s}^{5/2} r_{s}^5$ & $6.5\times 10^9\,L_\odot$\\
    $t_0$ & $(4\pi G \rho_{s})^{-1/2}$ & $51\,\text{Myr}$\\
    \hline
	\end{tabular}
\end{table}

We perform fluid simulations for the SIDM10-only, SIDM10+baryonM, D, and C benchmarks in Table~\ref{tab:benchmark}. In addition, we run fluid simulations with $\beta \sigma_m$ ranging from $0.075\, \text{cm}^2/\text{g}$ to $150\,\text{cm}^2/\text{g}$ for the baryonM, D, and C benchmarks, in order to investigate the universal evolution behavior of SIDM haloes with different cross sections and baryon distributions, as we will discuss in~\secref{scaling}. 

Our fluid simulations set the initial density profile of the dark matter halo to be an NFW profile, instead of a contracted CDM profile; see Fig.~\ref{fig:grow} (dashed vs. solid). This approach is justified as the halo evolution in the fluid model follows ``conduction-then-relaxation'' steps, i.e., the {\it entire} halo is relaxed to a state of hydrostatic equilibrium in the presence of the baryonic potential. Thus the influence of the baryons is dynamically incorporated in the fluid model itself. For the halo we consider, the thermalization timescale is ${\cal O}(0.01~{\rm Gyr})$ in the central region ($r\sim {\cal O}(0.1~{\rm kpc})$). Under such rapid thermalization, the halo properties are intensive to the initial setup after $\sim{\rm Gyr}$ of evolution, as we have demonstrated using N-body simulations in Fig.~\ref{fig:grow}. In practice, if we were to use a contracted density profile for the initial condition for the fluid simulations, the thermalization timescale would be slightly shortened. In this case, we may need to make a minor downward adjustment for $\beta$ to match with the N-body simulations, but the universal evolutionary behavior of the halo (as discussed in~\secref{scaling}) will remain the same.

\cite{Jiang:2022aqw} found that the utilization of a contracted density profile as a matching condition can improve the accuracy of the semi-analytical SIDM halo model proposed in~\cite{Kaplinghat:2013xca,Kaplinghat:2015aga}. In this model, the halo does not evolve dynamically, in contrast to the fluid model. In addition, the matching condition is imposed in the inner region, but the baryonic potential can affect the entire halo, causing the outer region to deviate from an NFW profile. Thus for the semi-analytical SIDM halo model, an explicit inclusion of the contraction effect is needed when the baryons are dynamically important.

\subsection{Quantities for characterizing gravothermal evolution of the halo}
\label{sec:keys}

From the N-body and fluid simulations, we can obtain the density and velocity dispersion profiles at different evolution times. From these snapshots, we can extract quantities that characterize the gravothermal evolution of the halo. Here, we discuss methods to obtain collapse time, time for maximal core expansion, central density, core size, and central velocity dispersion from the simulation snapshots.

\begin{enumerate}
\item  Collapse time $t_*$: We define $t_*$ as the elapsed time from $t=0$ until the onset of collapse when the Knudsen number, i.e., the ratio between the SIDM mean-free-path $\lambda$ and the local Jeans length $H$,
\beq
Kn \equiv \lambda/H = \sqrt{4\pi G \rho_\chi}/(\rho_\chi \sigma_m \nu_\chi)
\eeq 
reaches $0.1$ at the halo center (technically, the innermost layer of our fluid snapshots). In general, for $Kn<1$, a short-mean-free-path core forms in the collapsed central halo~\citep{Balberg:2001qg,Balberg:2002ue,Pollack:2014rja,Essig:2018pzq}. Since the central density grows rapidly in the deep collapse phase, adjusting the $Kn$ condition slightly does not affect the $t_*$ value. Although our N-body simulations cannot resolve the central region where $Kn<0.1$, we use them to calibrate the fluid model's $\beta$ parameter for each benchmark, then determine $t_*$ using the fluid simulations.

\item Central dark matter density $\rho_c$: For the N-body simulation, we evaluate the central halo density as the averaged density within $r < 2h$, where $h = 0.13\,\text{kpc}$ is the force softening length. For the fluid simulation, $\rho_c$ is computed as the density of the innermost layer of the fluid snapshots, i.e., the averaged density for the region with an enclosed mass of $4\pi \rho_{s} (10^{-2}r_s)^3$. Through the bulk of gravothermal evolution as we are interested in, the central density profile is rather flat within those radii. Hence, the difference between two ways of evaluating $r_c$ is minor. 

\item Dark matter density core size $r_c$: From the density profile of the halo, we compute the logarithmic density slope, $\d \log \rho_\chi/ \d \log r$. The core corresponds to the region where the slope is close to zero. To be concrete, we define the core size as the radius at which $\d \log \rho_\chi / \d \log r = -0.8$. Setting it to be a number closer to zero yields smaller $r_c$, but the trend of the evolution of $r_c$ remains the same. A complication arises when a short-mean-free-path core emerges on top of the collapsed central halo~\citep{Balberg:2002ue}, and we may get two values of the core size. When this occurs, we report the smaller value of the two.

\item Maximal core-expansion time $t_m$: $t_m$ can be evaluated as the moment when the central density $\rho_c$ reaches its minimum or when the core size $r_c$ reaches its maximum. For the fluid simulations, obtaining $t_m$ from the snapshots is straightforward because of their high resolution. For the N-body simulations, it could be difficult due to numerical fluctuations. To fix this issue, we fit the evolution curves of the central density and core size from the N-body simulations using the method of Gaussian process regression (GPR)~\citep[e.g.,][]{2022arXiv220908940A} and then determine $t_m$ using the fitted curves; see App.~\ref{sec:nbodyextract} for details.

\item Central 1D dark matter velocity dispersion $\nu_c$: We evaluate $\nu_c$ as the averaged velocity dispersion for dark matter particles within an averaging radius $r =0.4\,\text{kpc}$ from the halo center for the N-body simulations:
\beq
\nu_c^2 = \frac{1}{3}\langle \nu_{i, \text{3D}}^2\rangle_{r_i<0.4~{\rm kpc}},
\eeq 
where $\nu_{i, \text{3D}}$ is the 3D velocity dispersion for a particle and $i$ loops through all the particle within the radius. Averaging is necessary to suppress numerical fluctuations in the N-body simulations. Since $\nu_\chi$ is relatively flat in the central halo, reducing the averaging radius has little effect if the fluctuation can be ignored. We have also checked that increasing the averaging radius to $r=1~\text{kpc}$ only changes $\nu_c$ mildly. For the fluid simulations, we evaluate $\nu_c$ as 
\beq
\nu_c^2= \left.\frac{\int_0^{M_\chi} dM_\chi (r) \nu_\chi^2(r)}{M_\chi}\right|_{r=0.4~\text{kpc}}.
\eeq In practice, the integration is replaced by discrete summarization of the Lagrangian zones within $r=0.4~\textup{kpc}$.

\end{enumerate}

\section{Results}
\label{sec:results}

This section presents the results from our N-body and fluid simulations. We will mainly focus on the impact of the baryonic potential on the halo at different stages of gravothermal evolution, as well as its properties in the deep collapse phase. We further propose a simple formula for estimating the collapse time in the presence of the baryonic potential. 

\subsection{Accelerating core expansion and collapse}

\begin{figure}
	\includegraphics[width=0.993\columnwidth]{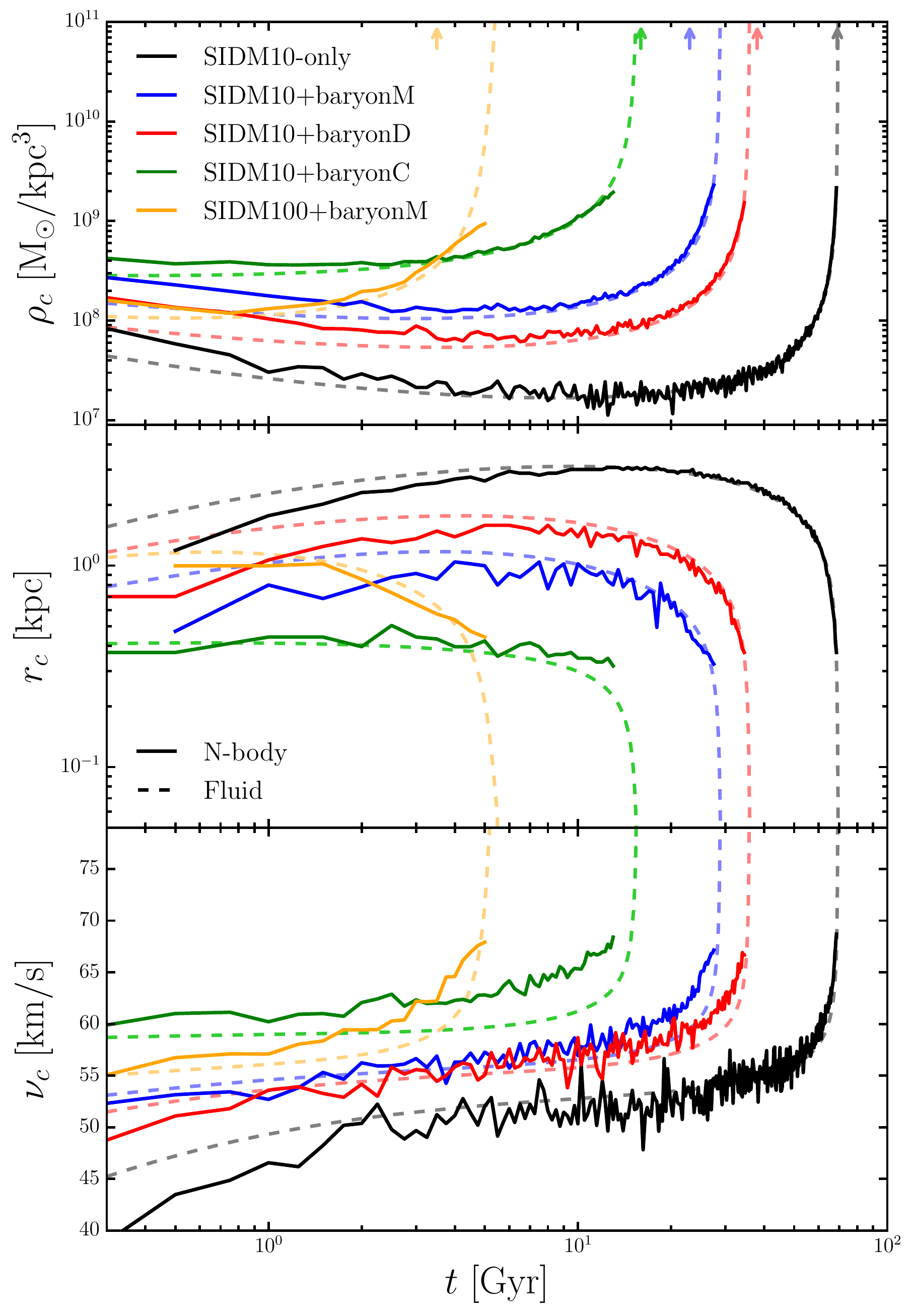}
    \caption{{\bf Top:} Evolution of the central dark matter density $\rho_c$ for the benchmarks with constant $\sigma_m$ listed in Table~\ref{tab:benchmark} from the N-body (solid) and fluid (dashed) simulations. For each benchmark, we allow the conduction coefficient $\beta$ to vary in the range $0.58\textup{--}0.91$ such that its fluid and N-body simulations agree when the collapse is substantial. The arrow denotes the collapse time estimated using~\eqref{estimate}. {\bf Middle:} Evolution of the dark matter core size $r_c$. {\bf Bottom:} Evolution of the central 1D dark matter velocity dispersion $\nu_c$. }
    \label{fig:densitytime1} 
\end{figure}

\begin{figure}
	~\includegraphics[width=\columnwidth]{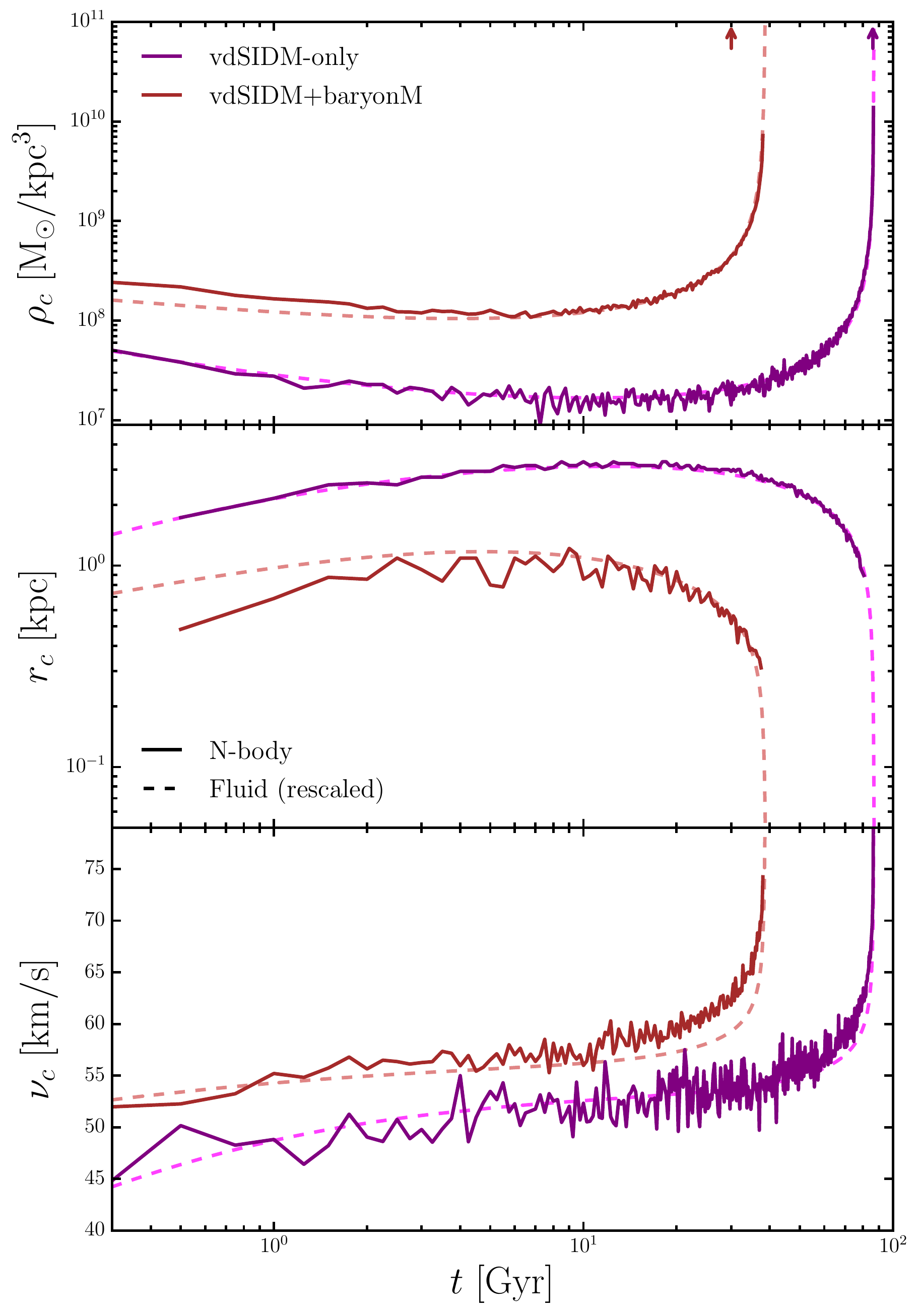}
    \caption{{\bf Top:} Evolution of the central dark matter density $\rho_c$, for the vdSIDM benchmarks from the N-body (solid) and fluid (dashed) simulations. The fluid simulations are performed based on the effective cross section $\sigma_m^\text{eff}$. The arrow denotes the collapse time estimated using~\eqref{estimate}. {\bf Middle:} Evolution of the dark matter core size $r_c$. {\bf Bottom:} Evolution of the central 1D dark matter velocity dispersion $\nu_c$.}
    \label{fig:densitytime2} 
\end{figure}

In~\figref{densitytime1}, we show the evolution of the central density $\rho_c$ (top), the core size $r_c$ (middle), and the central 1D velocity dispersion $\nu_c$ (bottom) for the five constant SIDM benchmarks, from the N-body (solid) and fluid (dashed) simulations. We have calibrated the fluid model by adjusting the conduction coefficient $\beta$ such that the model reproduces the evolution of $\rho_c$  from the N-body simulations at late stages. The resulting $\beta$ value is in a range of $0.58\textup{--}0.91$ for the constant SIDM benchmarks; see~Table~\ref{tab:extract}. 
We also find that once the fluid model is calibrated with $\rho_c$, it reasonably reproduces the evolution of $r_c$ and $\nu_c$ from the N-body simulations. \figref{densitytime2} shows the evolution of $\rho_c$ (top), $r_c$ (middle), and $\nu_c$ (bottom) for the velocity-dependent SIDM benchmarks from the N-body (solid) and fluid (dashed) simulations. We again see the agreement.  

The fluid model is calibrated for vdSIDM in the following way. We first follow~\cite{Yang:2022hkm} and calculate the effective constant cross sections as $\sigma^{\rm eff}_m=9.7\,\text{cm}^2/\text{g}$ and $8.6\,\text{cm}^2/\text{g}$ for the vdSIDM-only and vdSIDM+baryonM benchmarks, respectively. We have taken the effective 1D velocity dispersion to be $\nu_c(t_m)=53\,\text{km}/\text{s}$ for vdSIDM-only and $56\,\text{km}/\text{s}$ for vdSIDM+baryonM; see Table~\ref{tab:extract}. The mild increase in $\nu_c(t_m)$ due to the presence of the potential leads to the reduction of $\sigma_m^{\text{eff}}$ for vdSIDM+baryonM. We then rescale the SIDM10-only and SIDM10+baryonM fluid simulations with the calculated $\sigma^{\rm eff}_m$ values for vdSIDM-only and vdSIDM+baryonM, respectively, while adjusting $\beta$, such that the fluid model reproduces the evolution of $\rho_c$ from the N-body simulations. The calibrated $\beta$ values are reported in Table~\ref{tab:extract}.

\begin{table*}
	\centering
	\caption{Physical quantities characterizing the gravothermal evolution of the benchmark SIDM haloes (from left to right): labeling name, time for maximal core expansion $t_m$, central dark matter density at $t_m$ $\rho_{c}(t_m)$, 1D dark matter velocity dispersion $\nu_c(t_m)$, and dark matter core size $r_c(t_m)$,  collapse time $t_*$, calibrated conduction coefficient $\beta$, and estimated collapse time $t^\text{est}_*$ using~\eqref{estimate}. For the quantities related to $t_m$, we present two sets of values separated by a semicolon. The first set is extracted from the fluid simulations, and the second one is directly from the N-body simulations utilizing GPR, as shown in App.~\ref{sec:nbodyextract}. For the N-body simulations, $t_m$ values extracted from the $r_c(t)$ and $\rho_c(t)$ evolution trajectories could be different, and we report the relevant values separately if there is a noticeable difference: the one with a parenthesis $r_c (t)$, and the one without is from $\rho_c (t)$. }

	\label{tab:extract}

 	\begin{tabular}{l|ccc|ccccc|c|c} 
		\hline
		Name  &   $t_m $ [Gyr] & $\rho_{c}(t_m)$ [$10^7\text{M}_\odot/\text{kpc}^3$]  & $\nu_c(t_m)$ [km$/$s] & $r_c(t_m)$ [kpc] & $t_*$ [Gyr]  & $\beta$ & $t_*^\text{est}$ [Gyr]\\
		\hline
		SIDM10-only &  $8.4; 8.2(12)$  & $1.7;1.8$ & $53;53$ & $3.1;3.0$ & 69  & 0.84 & 69 \\
		SIDM10+baryonM  & $2.9;3.1(4.7)$ & $10;13$& $56;57$  & $1.2;1.0$ & 29 & 0.89 & 23 \\
		SIDM10+baryonD  & $3.6;6.2(4.9)$ & $5.4;6.6(7.0)$ & $55;56$ & $1.8;1.6$ & 36  & 0.91 & 38\\
		SIDM10+baryonC  & $0.37;1.8(2.4)$ & $28;36$ & $59;62$ & $0.41;0.44$ & 16 & 0.83  & 16\\
		\hline
		SIDM100+baryonM  & $0.52;0.72(1.1)$ & $11;13$ & $56;57$ & $1.2;1.0$ & 5.5 & 0.58  & 3.5\\
  		\hline
		vdSIDM-only   & $11;10(9.2)$ & $1.7;1.5$ & $53;53$ & $3.1;3.3$ & 87  & 0.69 & 86\\
		vdSIDM+baryonM & $3.9;8.6 (6.4)$ & $10;12$  & $56;57$ & $1.2;1.1$ & 39  & 0.78 & 30\\
		\hline
	\end{tabular}
\end{table*}

From Figs.~\ref{fig:densitytime1} and~\ref{fig:densitytime2}, we can obtain $t_*$, $t_m$, $\rho_c(t_m)$, $r_c(t_m)$, and $\nu_c(t_m)$ for the benchmarks, using the methods described in~\secref{keys}. The evaluation of $t_m$ needs further elaboration. For the fluid simulations, we search for the moment when $\rho_c(t)$ reaches its minimum or $r_c(t)$ reaches its maximum. The resulting $t_m$ values from the two searches coincide, and we read off $\nu_c(t_m)$ from the $\nu_c(t)$ evolution. For the N-body simulations, we first fit the simulated $\rho_c(t)$ and $r_c(t)$ data with GPR, see App.~\ref{sec:nbodyextract}, and then search for their minimal and maximum, respectively. The two searches do not necessarily yield the same $t_m$ value, and we report both in Table~\ref{tab:extract}, which could be regarded as a range where a true value of $t_m$ lies. The uncertainty in determining $t_m$ has minor effects on the evaluation of $\rho_c(t_m)$ and $r_c (t_m)$, as the halo properties near $t\sim t_{m}$ are relatively stable. Once the range of $t_m$ is specified, we choose an N-body snapshot within the range and identify $\nu_c (t_m)$; see Table~\ref{tab:extract}. We see that the N-body and fluid simulations agree well. 

\begin{figure}
	\includegraphics[width=\columnwidth]{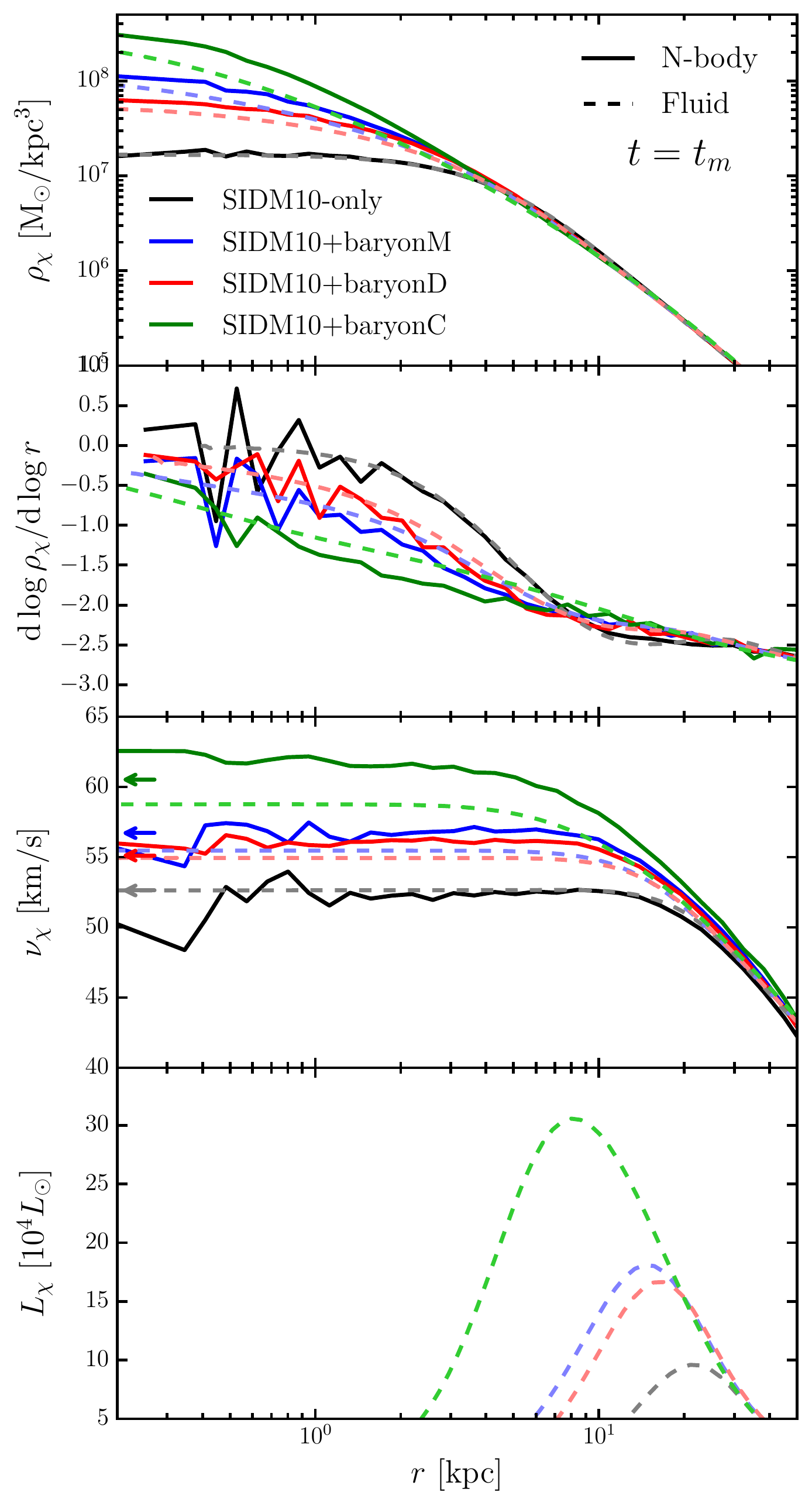}
    \caption{ {\bf From Top to Bottom:} Dark matter density, logarithmic density slope, velocity dispersion, and luminosity profiles, evaluated at the time of maximal core expansion $t=t_m$ for the SIDM10-only, SIDM10+baryonM, D, and C benchmarks from the N-body (solid) and fluid (dashed) simulations. The arrow in the third panel denotes the corresponding $\nu_c (0)$ value obtained by rescaling SIDM10-only $\nu_c(0)$ based on~\eqref{nuc_scale}. }

    \label{fig:key_snap_2} 
\end{figure}

In Fig.~\ref{fig:key_snap_2}, we further show profiles for the density, logarithmic density slope, velocity dispersion, and luminosity at $t=t_m$ from top to bottom panels. For the N-body simulations, we take the snapshots at $t_{m}\approx\{10,~4,~5.5,~2.25\}\,\text{Gyr}$ for SIDM10-only, SIDM10+baryonM, D, and C, respectively. The central velocity dispersion approximately follows the scaling relation
\beq
\nu_c(t_m) \propto \sqrt{|\Phi_c(t=0)|},
\label{eq:nuc_scale}
\eeq
In the third panel of Fig.~\ref{fig:key_snap_2}, the arrows denote the expected values of $\nu_c(t_m)$ for the benchmarks, using~\eqref{nuc_scale} and normalizing it for the SIDM10-only benchmark. 

The presence of the baryonic potential accelerates the gravothermal evolution of the halo and shorten the timescale for reaching the maximal core-expansion ($t_m$) and -collapse ($t_*$) phases, as well as their difference $t_*-t_m$. Following the trend, the core size decreases, and the velocity dispersion increases. The significance becomes higher as the baryon concentration increases. We also see that for $t=t_m$, the luminosity of all four benchmarks is positive everywhere in the halo, indicating the heat flow goes outward. Benchmarks with a deeper baryonic potential have a higher peak luminosity value and impact a broader range of radii. The increase in the positive luminosity persists throughout the gravothermal evolution. This enhancement leads to a substantial reduction in the collapse time, as, in its absence, developing a negative gradient in the velocity dispersion profile would be considerably longer.

For the SIDM10 benchmarks, the calibrated $\beta$ value varies in a small range $\beta=0.83\textup{--}0.91$, slightly higher than the canonical one $\beta=0.75$~\citep{Koda:2011yb}. On the other hand, for SIDM100+baryonM, $\beta=0.58$, which is lower. This difference could be because the central halo is close to the short-mean-free-path regime at $t=0$ for SIDM100; hence, its evolution is not entirely controlled by $\kappa_\text{lmfp}\propto \beta$. For the vdSIDM benchmarks, the calibrated $\beta$ values are $0.67$ and $0.78$ for vdSIDM-only and vdSIDM+baryonM, respectively. In addition, the fluid simulations tend to have faster core expansion than the N-body simulations (except for vdSIDM-only), although the agreement is excellent during the collapse phase. We find that the minor discrepancy could be fixed by multiplying the following time-dependent fudge factors to $\rho_c(t)$ and $r_c(t)$ from the fluid simulations,
\begin{equation}
f_{\rho_c}(t) = 1+e^{-25 t/ t_*},\quad f_{r_c} (t) = 1-0.4e^{-15 t/ t_*}
\label{eq:patch}
\end{equation}
respectively, for all the benchmarks.

\subsection{Universal halo properties in the deep collapse phase}

\begin{figure}
    \includegraphics[width=\columnwidth]{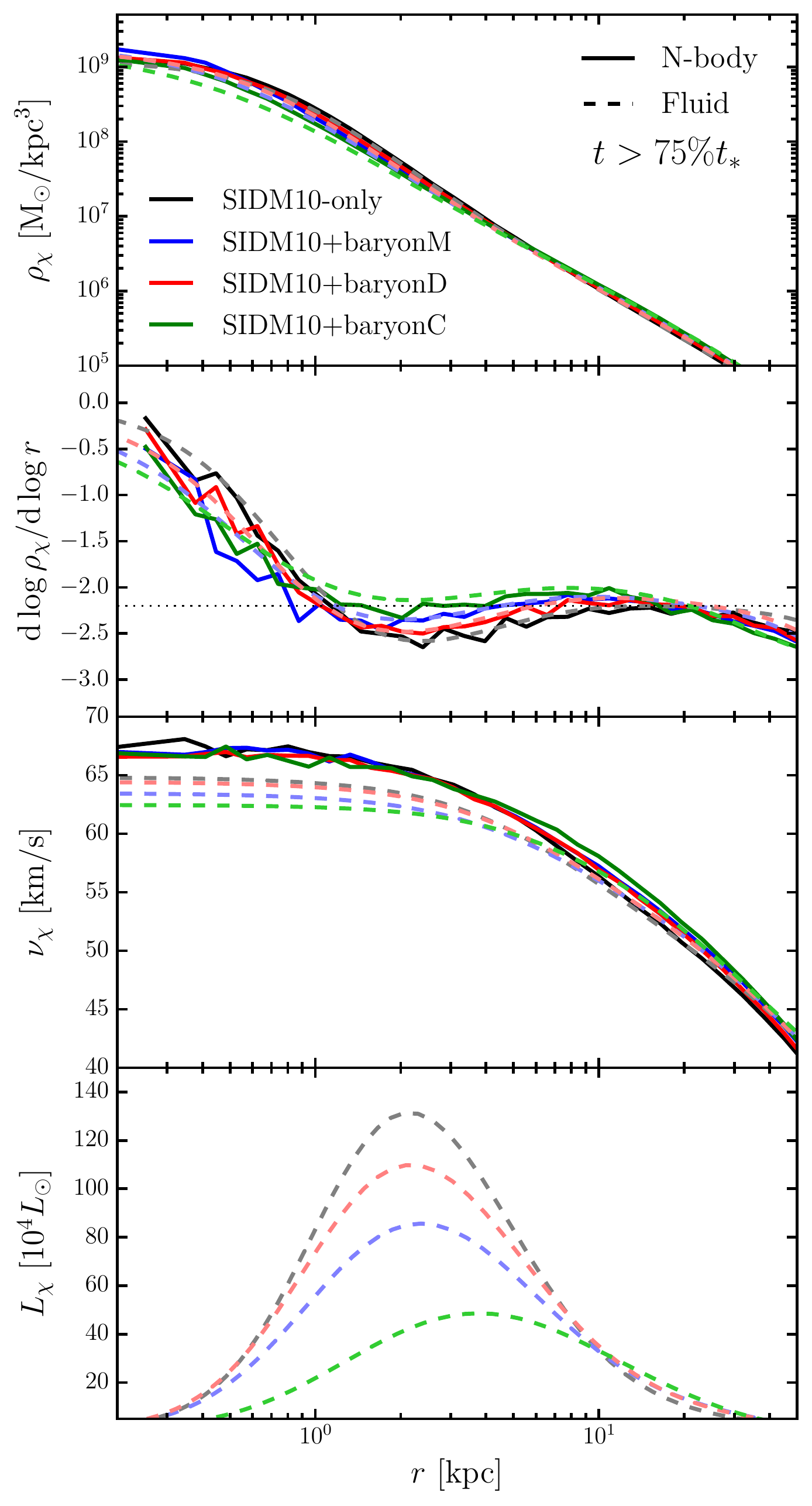}
    \caption{{\bf From Top to Bottom:} Dark matter density, logarithmic density slope, velocity dispersion, and luminosity profiles, evaluated at the time when the central densities are $10^9~{\rm M_\odot/kpc^3}$ for the SIDM10-only, SIDM10+baryonM, D, and C benchmarks from the N-body (solid) and fluid (dashed) simulations. In the second panel, the horizontal line indicates the value $\d \log \rho_\chi/\d \log r = -2.2$.}
    \label{fig:key_snap_3} 
\end{figure}

We check the halo properties in the deep collapse phase. For the four SIDM10 benchmarks, we chose snapshots such that their central densities are close to $10^{9}\,\text{M}_\odot/\text{kpc}^3$ and find $t=\{68~(98\% t_*),~27.25~(94\% t_*),~34.5~(96\% t_*),~12~(76\% t_*)\}~{\rm Gyr}$ for SIDM10-only, SIDM10+baryonM, D, and C, respectively. Fig.~\ref{fig:key_snap_3} shows the corresponding profiles of the density, logarithmic density slope, velocity dispersion, and luminosity from the N-body (solid) and fluid (dashed) simulations. These halo properties are similar for the four benchmarks (except for the luminosity profile). Such a universal behavior is consistent with what we expect from Fig.~\ref{fig:densitytime1}.
 
 Compared to the halo at $t=t_m$ shown in~\figref{key_snap_2}, the central density, velocity dispersion, and outward luminosity are significantly enhanced in the collapse phase. For $r= 1\textup{--}10\,\text{kpc}$, the density profile is cuspy, and its logarithmic slope is $-2.5 < \d \log \rho_\chi/\d \log r < -2$, with a slight tendency that faster collapse leads to a less cuspy profile. These results are broadly consistent with $\d \log \rho_\chi/\d \log r = -2.2$ from previous SIDM-only simulations~\citep[e.g.,][]{Koda:2011yb,Essig:2018pzq,Correa:2022dey,Outmezguine:2022bhq,Yang:2022hkm,Jiang:2022aqw}. The logarithmic slope asymptotes to $-0.5\textup{--}0$ for smaller radii, representing a collapsed central core.
  
The presence of the baryons can shorten the collapse timescale. Interestingly, once the central density of collapsed haloes is specified, the other properties do not depend on the baryon distribution explicitly. Thus we may use SIDM-only simulations to model the case with the baryons after rescaling the collapse time. This approach could be used in testing the gravothermal collapse of SIDM haloes with astrophysical observations, such as strong gravitational lensing systems~\citep{Yang:2021kdf,Minor:2020hic,Gilman:2021sdr, Gilman:2022ida,Loudas220913393,Nadler:2023nrd,Dhanasingham:2023thg}, supermassive black holes~\citep{Balberg:2001qg,Pollack:2014rja,Choquette:2018lvq,Feng:2020kxv,Feng:2021rst,Xiao:2021ftk,Meshveliani:2022rih}, and galactic rotation curves~\citep{Essig:2018pzq}.

\subsection{Estimating the collapse time}
\label{sec:estimate_collapse_time}

The significance of the baryonic potential in accelerating the collapse depends on its distribution. We propose a simple formula for estimating the collapse time in the presence of the baryons. For the SIDM-only case with a constant cross section, the collapse time can be parametrized as~\citep{Balberg:2002ue,Koda:2011yb,Essig:2018pzq} 
\beq
\label{eq:tc}
t_*   = \frac{150}{\beta \sigma_m} \frac{1}{ \rho_{s} r_{s}} \frac{1}{\sqrt{4\pi G \rho_{s}}}=\frac{150}{\beta\sigma_m} \frac{1}{ \rho_{s} \sqrt{|\Phi_{\chi}(0)|}|_{t=0}},
\eeq
where we have applied~\eqref{nfwpotential}, i.e., the gravitational potential of an NFW halo, for the last equality. We generalize it to our benchmarks with the following ansatz
\beq
t_*^\text{est} = \frac{150}{\beta \sigma_m^{\text{eff}}} \frac{1}{ \rho_{\rm eff} \sqrt{|\Phi_{\chi}(0)|+|\Phi_{b}(0)|}|_{t=0}},
\label{eq:estimate}
\eeq
where $\rho_{\rm eff}$ is an effective density that captures the contraction effect due to the baryonic potential. 
For NFW and Hernquist profiles, radius times density approaches $\rho_{s} r_{s}$ and $\rho_h r_h$ as $r$ goes to zero, respectively, and we evaluate $\rho_{\rm eff}$ as 
\beq
\rho_{\rm eff} = \frac{\rho_{s} r_{s} + \alpha \rho_h r_h}{r_{s} + \alpha r_h} \approx \rho_s + \frac{\alpha M_{b,\text{tot}}}{2\pi r_s r_h^2}, 
\eeq
where $\alpha$ is a weighting factor that parametrizes the relative significance of the baryonic potential. We find $\alpha\approx0.4$ works well for most of the benchmarks. In the last column of Table~\ref{tab:extract}, we list the collapse time estimated using~\eqref{estimate}, also denoted as the colored arrow in Figs.~\ref{fig:densitytime1} and~\ref{fig:densitytime2} (top). We see that agreement is better than $20\%$, expect for the extreme case SIDM100+baryonM. As discussed, with such a large cross section, the central halo is in the short-mean-free-path regime at $t\sim0$ and $\kappa\sim \kappa_{\rm smfp}\propto1/\sigma_m$. Thus we expect that for SIDM100, the collapse time estimated using~\eqref{estimate}, based on $\kappa\sim \kappa_{\rm lmfp}\propto\sigma_m$, is shorter than the actual one from the simulations. Such an underestimate is indeed the case; see~Table~\ref{tab:extract}.

\section{Universality of gravothermal evolution}
\label{sec:scaling}

Studies show that the gravothermal evolution of SIDM haloes exhibits a quasi-universal behavior, e.g., after appropriately rescaling, the evolution of the halo properties are almost independent of a particular choice of initial halo parameters $r_s$ and $\rho_s$~\citep{Balberg:2002ue,Pollack:2014rja,Essig:2018pzq}, as well as $\sigma_m$~\citep{Outmezguine:2022bhq,Yang:2023jwn}. These studies are based on SIDM-only fluid simulations. We investigate the universality of the SIDM haloes in the presence of the baryonic potential for the following two scenarios: fixing the baryonic potential while varying the cross section strength; further allowing the baryonic potential to be different.

\begin{figure}
    \centering
    \includegraphics[width=0.98\columnwidth]{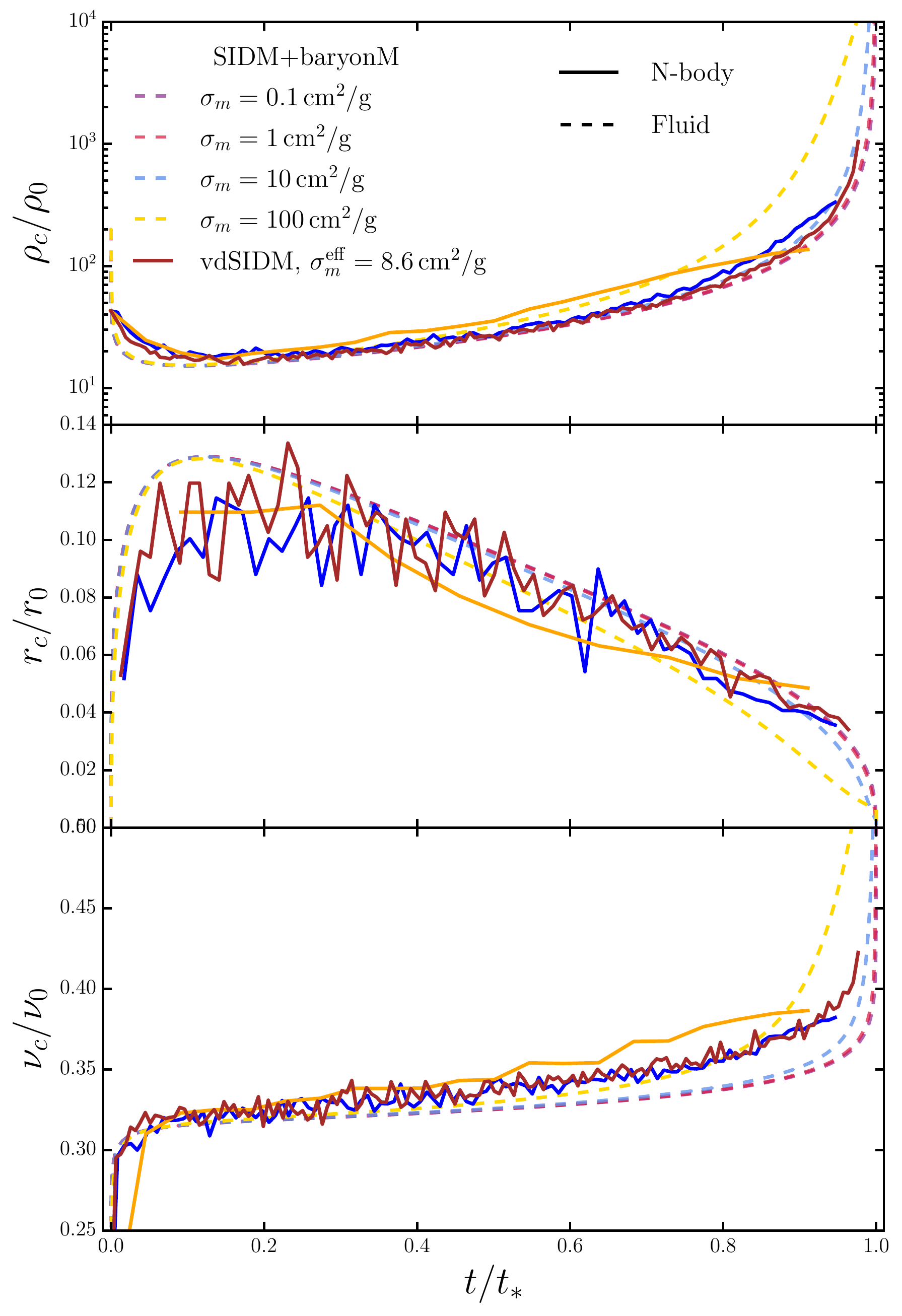}
    \caption{{\bf Top:} Evolution of the central dark matter density for the (vd)SIDM+baryonM benchmarks from the fluid simulations with constant $\sigma_m = 0.1, 1, 10$, and $100\,\text{cm}^2/\text{g}$ (dashed), and the N-body simulations with constant $\sigma_m = 10$ and $100\,\text{cm}^2/\text{g}$, as well as vdSIDM+baryonM (solid). The density and evolution time are normalized as $\rho/\rho_0$ and $t/t_*$, respectively. {\bf Middle:} Evolution of the core size, normalized to $r_c/r_0$. {\bf Bottom:} Evolution of the central 1D velocity dispersion, normalized to $\nu_c/\nu_0$. }
    \label{fig:universality1}
\end{figure}

\figref{universality1} shows the evolution of normalized $\rho_c$, $r_c$, and $\nu_c$ for the SIDM+baryonM initial condition with constant cross sections $\sigma_m=0.1,1,10$ and $100\,\text{cm}^2/\text{g}$ from the fluid simulations (dashed), after applying the rescaling relation
\begin{align}
t \to \frac{t}{t_*}, \quad \rho_{c} \to \frac{\rho_{c}}{\rho_0},\quad \nu_{c}\to \frac{\nu_{c}}{\nu_0}, \quad r_{c} \to \frac{r_c}{r_0}.
\label{eq:scaling}
\end{align}
The evolution trajectories for $\sigma_m = 0.1,1$, and $10\,\text{cm}^2/\text{g}$ mostly overlap, manifesting the universal behavior. The $\sigma_m=100\,\text{cm}^2/\text{g}$ case exhibits a similar trend, although the deviation becomes significant in the deep collapse phase. For SIDM10+baryonM and SIDM100+baryonM, as well as vdSIDM+baryonM, we also show their N-body simulations (solid). We have further checked that the universal behavior holds for the SIDM-only, SIDM+baryonD, and C benchmarks.

\begin{figure}
    \centering
    ~\includegraphics[width=0.97\columnwidth]{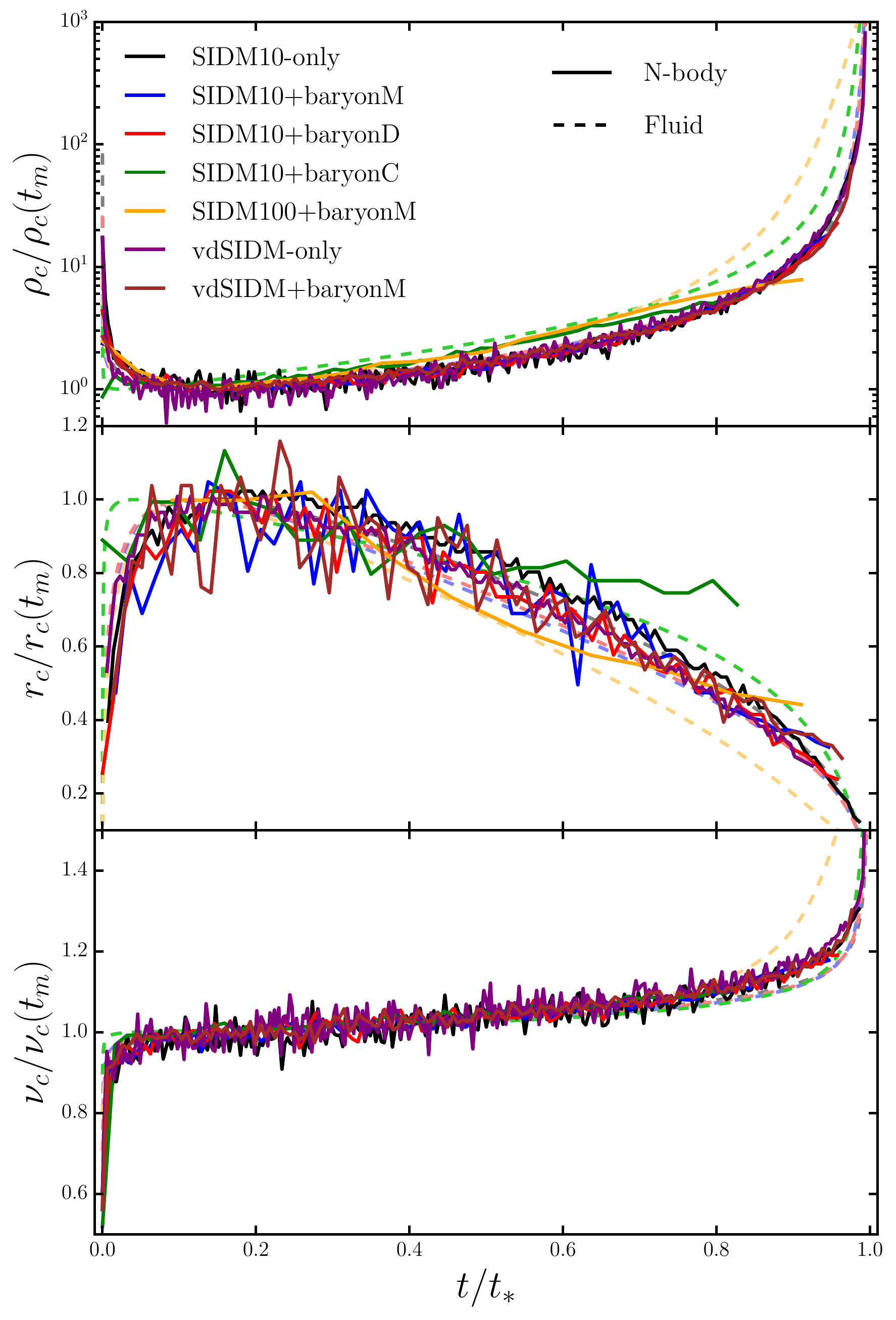}
    \caption{{\bf Top:} Evolution of the central dark matter density from the N-body (solid) and fluid (dashed) simulations for all benchmarks listed in Table~\ref{tab:benchmark}.  For each benchmark, the evolution time and the central density are normalized to $t/t_*$, and to minima, i.e.,  the value at the maximal core expansion, $\rho_c(t_m)$, respectively. {\bf Middle:}  Evolution of the dark matter core size $r_c$, normalized to $r_c/r_c(t_m)$.  {\bf Bottom:} Evolution of the central 1D velocity dispersion, normalized to $\nu_c/\nu_c(t_m)$.
   }
    \label{fig:universality2}
\end{figure}

We can understand the universal behavior based on the fluid model. The evolution equation in~\eqref{fluiddimless} is already dimensionless but still depends on $\hat\sigma_m$.  We assume the bulk of the evolution is in the long-mean-free-path regime and write the collapse time is in the form ${t_*}/{t_0} = {\gamma}/{\beta \hat \sigma_m}$, where $\gamma$ is a constant. Then we rescale the evolution time and luminosity as $\hat{\hat t} \equiv {t}/{t_*} = (C \hat \sigma_m/\gamma)(t/t_0) $ and $\hat{\hat L}_\chi \equiv (\gamma/C\hat \sigma_m) (L/L_0)$, and express~\eqref{fluiddimless} as 
\begin{align}
\frac{\partial \hat M_{\chi}}{\partial \hat r} ={}&\hat r^2 \hat \rho_\chi,\quad \frac{\partial (\hat \rho_\chi \hat \nu_\chi^2)}{\partial \hat r} = -\frac{(\hat M_\chi + \hat M_b)\hat \rho_\chi}{\hat r^2}, \nonumber\\
\frac{\partial \hat{\hat L}_\chi}{\partial \hat r} ={}& -\hat \rho_\chi \hat r^2\hat \nu_\chi^2 D_{\hat{\hat t}} \ln \frac{\hat \nu_\chi^3}{\hat \rho_\chi}, \quad 
\frac{\hat{\hat L}_\chi}{\hat r^2} = -3.4 \gamma \hat \rho_\chi \hat \nu_\chi^3 \frac{\partial \hat \nu_\chi^2}{\partial \hat r},
\label{eq:fluiddimless2}
\end{align}
where we have assumed $\kappa\approx\kappa_{\rm lmfp}$, and it is valid for $\hat{\sigma}_m<1$. \eqref{fluiddimless2} does not depend on $\hat \sigma$ explicitly. Thus under the rescaling relation \eqref{scaling}, the halo evolution with a baryonic potential, but different $\sigma_m$ values, exhibit the university. For $\sigma_m=100~{\rm cm^2/g}$, the deviation in the deep collapse phase is likely due to the fact central halo evolves into the short-mean-free-path regime, where the assumption $\kappa\approx\kappa_{\rm lmfp}$ breaks down.

The relation in~\eqref{scaling} does not eliminate the explicit dependence on the baryonic potential ($\hat M_b$). Intriguingly, we find that after applying the following rescaling relation
\beq
t \to \frac{t}{t_*},~\rho_c \to \frac{\rho_c}{\rho_c(t_m)},~\nu_c \to \frac{\nu_c}{\nu_c(t_m)},~r_c\to \frac{r_c}{r_c(t_m)},
\label{eq:scaling2}
\eeq
the dependence on the potential becomes implicit, and all benchmarks we consider evolve universally, as demonstrated in~\figref{universality2}. The specific values of $\rho_c(t_m)$, $\nu_c(t_m)$, and $r_c(t_m)$ are adopted from~Table~\ref{tab:extract}. For the N-body simulations, if the corresponding values extracted from $\rho_c(t)$ and $r_c(t)$ are different, we take an average of the two.

\section{Conclusions}
\label{sec:conclusion}
In this work, we have used the controlled N-body and fluid simulations to study the impact of a baryonic potential on the gravothermal evolution of SIDM haloes. The presence of the potential can shorten the timescale for the halo to reach the maximal core-expansion and -collapse phases, and the significance is correlated with the concentration of the baryons. We explicitly showed that the final SIDM halo properties are robust to the formation history of the potential due to SIDM thermalization. We extended the fluid model to incorporate the effect of the baryonic potential and calibrated it with our N-body simulations. The  calibrated model successfully predicts the evolution of the halo properties, such as, the central density, core size, and velocity dispersion of dark matter particles. 

We further showed that even in the presence of the baryons, the evolution of SIDM haloes exhibits universality, a feature previously known for the SIDM-only case. For a fixed baryonic potential, the explicit dependence on the cross section can be absorbed by rescaling the evolution time with the collapse time, similar to the SIDM-only case. More interestingly, we introduced a new set of fiducial quantities under which the evolution of rescaled central density, velocity dispersion, and core size becomes universal, although the baryon distributions are different. The universality can be violated if the cross section is too large and the central halo is in the short-mean-free-path regime. 

Our simulations are based on an idealized setup, and it would be interesting to extend the study to hydrodynamical SIDM simulations of galaxy formation. As an example, we can construct a fluid model that incorporates a time-varying baryonic potential and calibrate it using hydrodynamical simulations. In particular, the time dependence of the mass and size of the baryon component can be directly obtained from those simulations. We can also study if the evolution of SIDM haloes in cosmological environments also exhibits universality, considering the fact that both halo and baryon mass change with time. We leave them for future work.

\section*{Acknowledgements}

We thank Stuart Shapiro, Shengqi Yang, and Moritz Fischer for helpful discussions. The fluid simulations were performed at the University of Chicago's Research Computing Center. We thank Edward W. Kolb for providing access to the resources. YZ acknowledges the Aspen Center for Physics for its hospitality during the completion of this study, which is supported by the National Science Foundation under Grant PHY-1607611. YZ was partially supported by the Kavli Institute for Cosmological Physics at the University of Chicago through an endowment from the Kavli Foundation and its founder Fred Kavli and partially supported by grants from the City University of Hong Kong (Project No. 9610645). DY and HBY were supported by the US Department of Energy under Grant DE-SC0008541 and the John Templeton Foundation under Grant 61884. The opinions expressed in this publication are those of the authors and do not necessarily reflect the views of the agencies.

\section*{Data Availability}

The simulation data can be obtained by making a reasonable request to the authors.


\bibliographystyle{mnras}
\bibliography{paper} 

\begin{thebibliography}{}
\makeatletter
\relax
\def\mn@urlcharsother{\let\do\@makeother \do\$\do\&\do\#\do\^\do\_\do\%\do\~}
\def\mn@doi{\begingroup\mn@urlcharsother \@ifnextchar [ {\mn@doi@}
  {\mn@doi@[]}}
\def\mn@doi@[#1]#2{\def\@tempa{#1}\ifx\@tempa\@empty \href
  {http://dx.doi.org/#2} {doi:#2}\else \href {http://dx.doi.org/#2} {#1}\fi
  \endgroup}
\def\mn@eprint#1#2{\mn@eprint@#1:#2::\@nil}
\def\mn@eprint@arXiv#1{\href {http://arxiv.org/abs/#1} {{\tt arXiv:#1}}}
\def\mn@eprint@dblp#1{\href {http://dblp.uni-trier.de/rec/bibtex/#1.xml}
  {dblp:#1}}
\def\mn@eprint@#1:#2:#3:#4\@nil{\def\@tempa {#1}\def\@tempb {#2}\def\@tempc
  {#3}\ifx \@tempc \@empty \let \@tempc \@tempb \let \@tempb \@tempa \fi \ifx
  \@tempb \@empty \def\@tempb {arXiv}\fi \@ifundefined
  {mn@eprint@\@tempb}{\@tempb:\@tempc}{\expandafter \expandafter \csname
  mn@eprint@\@tempb\endcsname \expandafter{\@tempc}}}

\bibitem[\protect\citeauthoryear{{Adhikari} et~al.,}{{Adhikari}
  et~al.}{2022}]{2022arXiv220710638A}
{Adhikari} S.,  et~al., 2022, \mn@doi [arXiv e-prints]
  {10.48550/arXiv.2207.10638}, \href
  {https://ui.adsabs.harvard.edu/abs/2022arXiv220710638A} {p. arXiv:2207.10638}

\bibitem[\protect\citeauthoryear{{Aigrain} \& {Foreman-Mackey}}{{Aigrain} \&
  {Foreman-Mackey}}{2022}]{2022arXiv220908940A}
{Aigrain} S.,  {Foreman-Mackey} D.,  2022, arXiv e-prints, \href
  {https://ui.adsabs.harvard.edu/abs/2022arXiv220908940A} {p. arXiv:2209.08940}

\bibitem[\protect\citeauthoryear{Andrade, Fuson, Gad-Nasr, Kong, Minor, Roberts
   \& Kaplinghat}{Andrade et~al.}{2021}]{Andrade:2020lqq}
Andrade K.~E.,  Fuson J.,  Gad-Nasr S.,  Kong D.,  Minor Q.,  Roberts M.~G.,
  Kaplinghat M.,  2021, \mn@doi [Mon. Not. Roy. Astron. Soc.]
  {10.1093/mnras/stab3241}, 510, 54

\bibitem[\protect\citeauthoryear{Balberg \& Shapiro}{Balberg \&
  Shapiro}{2002}]{Balberg:2001qg}
Balberg S.,  Shapiro S.~L.,  2002, \mn@doi [Phys. Rev. Lett.]
  {10.1103/PhysRevLett.88.101301}, 88, 101301

\bibitem[\protect\citeauthoryear{Balberg, Shapiro  \& Inagaki}{Balberg
  et~al.}{2002}]{Balberg:2002ue}
Balberg S.,  Shapiro S.~L.,   Inagaki S.,  2002, \mn@doi [Astrophys. J.]
  {10.1086/339038}, 568, 475

\bibitem[\protect\citeauthoryear{Blumenthal, Faber, Flores  \&
  Primack}{Blumenthal et~al.}{1986}]{Blumenthal:1985qy}
Blumenthal G.~R.,  Faber S.~M.,  Flores R.,   Primack J.~R.,  1986, \mn@doi
  [Astrophys. J.] {10.1086/163867}, 301, 27

\bibitem[\protect\citeauthoryear{Burger, Zavala, Sales, Vogelsberger, Marinacci
   \& Torrey}{Burger et~al.}{2022}]{Burger:2022cjo}
Burger J.~D.,  Zavala J.,  Sales L.~V.,  Vogelsberger M.,  Marinacci F.,
  Torrey P.,  2022, \mn@doi [Mon. Not. Roy. Astron. Soc.]
  {10.1093/mnras/stac994}, 513, 3458

\bibitem[\protect\citeauthoryear{{Carleton}, {Errani}, {Cooper}, {Kaplinghat},
  {Pe{\~n}arrubia}  \& {Guo}}{{Carleton} et~al.}{2019}]{2019MNRAS.485..382C}
{Carleton} T.,  {Errani} R.,  {Cooper} M.,  {Kaplinghat} M.,  {Pe{\~n}arrubia}
  J.,   {Guo} Y.,  2019, \mn@doi [\mnras] {10.1093/mnras/stz383}, \href
  {https://ui.adsabs.harvard.edu/abs/2019MNRAS.485..382C} {485, 382}

\bibitem[\protect\citeauthoryear{Choquette, Cline  \& Cornell}{Choquette
  et~al.}{2019}]{Choquette:2018lvq}
Choquette J.,  Cline J.~M.,   Cornell J.~M.,  2019, \mn@doi [JCAP]
  {10.1088/1475-7516/2019/07/036}, 07, 036

\bibitem[\protect\citeauthoryear{Correa}{Correa}{2021}]{Correa:2020qam}
Correa C.~A.,  2021, \mn@doi [Mon. Not. Roy. Astron. Soc.]
  {10.1093/mnras/stab506}, 503, 920

\bibitem[\protect\citeauthoryear{{Correa}, {Schaller}, {Ploeckinger}, {Anau
  Montel}, {Weniger}  \& {Ando}}{{Correa} et~al.}{2022}]{Correa:2022dey}
{Correa} C.~A.,  {Schaller} M.,  {Ploeckinger} S.,  {Anau Montel} N.,
  {Weniger} C.,   {Ando} S.,  2022, \mn@doi [\mnras] {10.1093/mnras/stac2830},
  \href {https://ui.adsabs.harvard.edu/abs/2022MNRAS.517.3045C} {517, 3045}

\bibitem[\protect\citeauthoryear{Creasey, Sameie, Sales, Yu, Vogelsberger  \&
  Zavala}{Creasey et~al.}{2017}]{Creasey:2017qxc}
Creasey P.,  Sameie O.,  Sales L.~V.,  Yu H.-B.,  Vogelsberger M.,   Zavala J.,
   2017, \mn@doi [Mon. Not. Roy. Astron. Soc.] {10.1093/mnras/stx522}, 468,
  2283

\bibitem[\protect\citeauthoryear{Dave, Spergel, Steinhardt  \& Wandelt}{Dave
  et~al.}{2001}]{Dave:2000ar}
Dave R.,  Spergel D.~N.,  Steinhardt P.~J.,   Wandelt B.~D.,  2001, \mn@doi
  [Astrophys. J.] {10.1086/318417}, 547, 574

\bibitem[\protect\citeauthoryear{Despali, Sparre, Vegetti, Vogelsberger, Zavala
   \& Marinacci}{Despali et~al.}{2019}]{Despali:2018zpw}
Despali G.,  Sparre M.,  Vegetti S.,  Vogelsberger M.,  Zavala J.,   Marinacci
  F.,  2019, \mn@doi [Mon. Not. Roy. Astron. Soc.] {10.1093/mnras/stz273}, 484,
  4563

\bibitem[\protect\citeauthoryear{Dhanasingham, Cyr-Racine, Mace, Peter  \&
  Benson}{Dhanasingham et~al.}{2023}]{Dhanasingham:2023thg}
Dhanasingham B.,  Cyr-Racine F.-Y.,  Mace C.,  Peter A. H.~G.,   Benson A.,
  2023

\bibitem[\protect\citeauthoryear{Dutton \& Macci\`o}{Dutton \&
  Macci\`o}{2014}]{Dutton:2014xda}
Dutton A.~A.,  Macci\`o A.~V.,  2014, \mn@doi [Mon. Not. Roy. Astron. Soc.]
  {10.1093/mnras/stu742}, 441, 3359

\bibitem[\protect\citeauthoryear{Elbert, Bullock, Kaplinghat, Garrison-Kimmel,
  Graus  \& Rocha}{Elbert et~al.}{2018}]{Elbert:2016dbb}
Elbert O.~D.,  Bullock J.~S.,  Kaplinghat M.,  Garrison-Kimmel S.,  Graus
  A.~S.,   Rocha M.,  2018, \mn@doi [Astrophys. J.] {10.3847/1538-4357/aa9710},
  853, 109

\bibitem[\protect\citeauthoryear{Essig, Mcdermott, Yu  \& Zhong}{Essig
  et~al.}{2019}]{Essig:2018pzq}
Essig R.,  Mcdermott S.~D.,  Yu H.-B.,   Zhong Y.-M.,  2019, \mn@doi [Phys.
  Rev. Lett.] {10.1103/PhysRevLett.123.121102}, 123, 121102

\bibitem[\protect\citeauthoryear{Feng, Kaplinghat, Tu  \& Yu}{Feng
  et~al.}{2009}]{feng:2009mn}
Feng J.~L.,  Kaplinghat M.,  Tu H.,   Yu H.-B.,  2009, \mn@doi [JCAP]
  {10.1088/1475-7516/2009/07/004}, 07, 004

\bibitem[\protect\citeauthoryear{Feng, Yu  \& Zhong}{Feng
  et~al.}{2021}]{Feng:2020kxv}
Feng W.-X.,  Yu H.-B.,   Zhong Y.-M.,  2021, \mn@doi [Astrophys. J. Lett.]
  {10.3847/2041-8213/ac04b0}, 914, L26

\bibitem[\protect\citeauthoryear{Feng, Yu  \& Zhong}{Feng
  et~al.}{2022}]{Feng:2021rst}
Feng W.-X.,  Yu H.-B.,   Zhong Y.-M.,  2022, \mn@doi [JCAP]
  {10.1088/1475-7516/2022/05/036}, 05, 036

\bibitem[\protect\citeauthoryear{Fischer, Br\"uggen, Schmidt-Hoberg, Dolag,
  Kahlhoefer, Ragagnin  \& Robertson}{Fischer et~al.}{2022}]{Fischer:2022rko}
Fischer M.~S.,  Br\"uggen M.,  Schmidt-Hoberg K.,  Dolag K.,  Kahlhoefer F.,
  Ragagnin A.,   Robertson A.,  2022, \mn@doi [Mon. Not. Roy. Astron. Soc.]
  {10.1093/mnras/stac2207}, 516, 1923

\bibitem[\protect\citeauthoryear{Gilman, Bovy, Treu, Nierenberg, Birrer, Benson
   \& Sameie}{Gilman et~al.}{2021}]{Gilman:2021sdr}
Gilman D.,  Bovy J.,  Treu T.,  Nierenberg A.,  Birrer S.,  Benson A.,   Sameie
  O.,  2021, \mn@doi [Mon. Not. Roy. Astron. Soc.] {10.1093/mnras/stab2335},
  507, 2432

\bibitem[\protect\citeauthoryear{Gilman, Zhong  \& Bovy}{Gilman
  et~al.}{2023}]{Gilman:2022ida}
Gilman D.,  Zhong Y.-M.,   Bovy J.,  2023, \mn@doi [Phys. Rev. D]
  {10.1103/PhysRevD.107.103008}, 107, 103008

\bibitem[\protect\citeauthoryear{{Hernquist}}{{Hernquist}}{1990}]{1990ApJ...356..359H}
{Hernquist} L.,  1990, \mn@doi [\apj] {10.1086/168845}, \href
  {https://ui.adsabs.harvard.edu/abs/1990ApJ...356..359H} {356, 359}

\bibitem[\protect\citeauthoryear{Huo, Yu  \& Zhong}{Huo
  et~al.}{2020}]{Huo:2019yhk}
Huo R.,  Yu H.-B.,   Zhong Y.-M.,  2020, \mn@doi [JCAP]
  {10.1088/1475-7516/2020/06/051}, 06, 051

\bibitem[\protect\citeauthoryear{Ibe \& Yu}{Ibe \& Yu}{2010}]{Ibe:2009mk}
Ibe M.,  Yu H.-b.,  2010, \mn@doi [Phys. Lett. B]
  {10.1016/j.physletb.2010.07.026}, 692, 70

\bibitem[\protect\citeauthoryear{{Jiang} et~al.,}{{Jiang}
  et~al.}{2023}]{Jiang:2022aqw}
{Jiang} F.,  et~al., 2023, \mn@doi [\mnras] {10.1093/mnras/stad705}, \href
  {https://ui.adsabs.harvard.edu/abs/2023MNRAS.521.4630J} {521, 4630}

\bibitem[\protect\citeauthoryear{Kahlhoefer, Kaplinghat, Slatyer  \&
  Wu}{Kahlhoefer et~al.}{2019}]{Kahlhoefer:2019oyt}
Kahlhoefer F.,  Kaplinghat M.,  Slatyer T.~R.,   Wu C.-L.,  2019, \mn@doi
  [JCAP] {10.1088/1475-7516/2019/12/010}, 12, 010

\bibitem[\protect\citeauthoryear{Kamada, Kaplinghat, Pace  \& Yu}{Kamada
  et~al.}{2017}]{Kamada:2016euw}
Kamada A.,  Kaplinghat M.,  Pace A.~B.,   Yu H.-B.,  2017, \mn@doi [Phys. Rev.
  Lett.] {10.1103/PhysRevLett.119.111102}, 119, 111102

\bibitem[\protect\citeauthoryear{Kaplinghat, Keeley, Linden  \& Yu}{Kaplinghat
  et~al.}{2014}]{Kaplinghat:2013xca}
Kaplinghat M.,  Keeley R.~E.,  Linden T.,   Yu H.-B.,  2014, \mn@doi [Phys.
  Rev. Lett.] {10.1103/PhysRevLett.113.021302}, 113, 021302

\bibitem[\protect\citeauthoryear{Kaplinghat, Tulin  \& Yu}{Kaplinghat
  et~al.}{2016}]{Kaplinghat:2015aga}
Kaplinghat M.,  Tulin S.,   Yu H.-B.,  2016, \mn@doi [Phys. Rev. Lett.]
  {10.1103/PhysRevLett.116.041302}, 116, 041302

\bibitem[\protect\citeauthoryear{Kaplinghat, Valli  \& Yu}{Kaplinghat
  et~al.}{2019}]{Kaplinghat:2019svz}
Kaplinghat M.,  Valli M.,   Yu H.-B.,  2019, \mn@doi [Mon. Not. Roy. Astron.
  Soc.] {10.1093/mnras/stz2511}, 490, 231

\bibitem[\protect\citeauthoryear{Koda \& Shapiro}{Koda \&
  Shapiro}{2011}]{Koda:2011yb}
Koda J.,  Shapiro P.~R.,  2011, \mn@doi [Mon. Not. Roy. Astron. Soc.]
  {10.1111/j.1365-2966.2011.18684.x}, 415, 1125

\bibitem[\protect\citeauthoryear{{Loudas}, {Pavlidou}, {Casadio}  \&
  {Tassis}}{{Loudas} et~al.}{2022}]{Loudas220913393}
{Loudas} N.,  {Pavlidou} V.,  {Casadio} C.,   {Tassis} K.,  2022, \mn@doi
  [\aap] {10.1051/0004-6361/202244978}, \href
  {https://ui.adsabs.harvard.edu/abs/2022A&A...668A.166L} {668, A166}

\bibitem[\protect\citeauthoryear{Meshveliani, Zavala  \& Lovell}{Meshveliani
  et~al.}{2023}]{Meshveliani:2022rih}
Meshveliani T.,  Zavala J.,   Lovell M.~R.,  2023, \mn@doi [Phys. Rev. D]
  {10.1103/PhysRevD.107.083010}, 107, 083010

\bibitem[\protect\citeauthoryear{Minor, Gad-Nasr, Kaplinghat  \& Vegetti}{Minor
  et~al.}{2021}]{Minor:2020hic}
Minor Q.~E.,  Gad-Nasr S.,  Kaplinghat M.,   Vegetti S.,  2021, \mn@doi [Mon.
  Not. Roy. Astron. Soc.] {10.1093/mnras/stab2247}, 507, 1662

\bibitem[\protect\citeauthoryear{{Moster}, {Naab}  \& {White}}{{Moster}
  et~al.}{2013}]{2013MNRAS.428.3121M}
{Moster} B.~P.,  {Naab} T.,   {White} S. D.~M.,  2013, \mn@doi [\mnras]
  {10.1093/mnras/sts261}, \href
  {https://ui.adsabs.harvard.edu/abs/2013MNRAS.428.3121M} {428, 3121}

\bibitem[\protect\citeauthoryear{Nadler, Banerjee, Adhikari, Mao  \&
  Wechsler}{Nadler et~al.}{2020}]{Nadler:2020ulu}
Nadler E.~O.,  Banerjee A.,  Adhikari S.,  Mao Y.-Y.,   Wechsler R.~H.,  2020,
  \mn@doi [Astrophys. J.] {10.3847/1538-4357/ab94b0}, 896, 112

\bibitem[\protect\citeauthoryear{{Nadler}, {Yang}  \& {Yu}}{{Nadler}
  et~al.}{2023}]{Nadler:2023nrd}
{Nadler} E.~O.,  {Yang} D.,   {Yu} H.-B.,  2023, \mn@doi [arXiv e-prints]
  {10.48550/arXiv.2306.01830}, \href
  {https://ui.adsabs.harvard.edu/abs/2023arXiv230601830N} {p. arXiv:2306.01830}

\bibitem[\protect\citeauthoryear{Navarro, Frenk  \& White}{Navarro
  et~al.}{1997}]{Navarro:1996gj}
Navarro J.~F.,  Frenk C.~S.,   White S. D.~M.,  1997, \mn@doi [Astrophys. J.]
  {10.1086/304888}, 490, 493

\bibitem[\protect\citeauthoryear{Nishikawa, Boddy  \& Kaplinghat}{Nishikawa
  et~al.}{2020}]{Nishikawa:2019lsc}
Nishikawa H.,  Boddy K.~K.,   Kaplinghat M.,  2020, \mn@doi [Phys. Rev. D]
  {10.1103/PhysRevD.101.063009}, 101, 063009

\bibitem[\protect\citeauthoryear{{Outmezguine}, {Boddy}, {Gad-Nasr},
  {Kaplinghat}  \& {Sagunski}}{{Outmezguine}
  et~al.}{2022}]{Outmezguine:2022bhq}
{Outmezguine} N.~J.,  {Boddy} K.~K.,  {Gad-Nasr} S.,  {Kaplinghat} M.,
  {Sagunski} L.,  2022, \mn@doi [arXiv e-prints] {10.48550/arXiv.2204.06568},
  \href {https://ui.adsabs.harvard.edu/abs/2022arXiv220406568O} {p.
  arXiv:2204.06568}

\bibitem[\protect\citeauthoryear{Pollack, Spergel  \& Steinhardt}{Pollack
  et~al.}{2015}]{Pollack:2014rja}
Pollack J.,  Spergel D.~N.,   Steinhardt P.~J.,  2015, \mn@doi [\apj]
  {10.1088/0004-637X/804/2/131}, 804, 131

\bibitem[\protect\citeauthoryear{Rahimi, Vienneau, Bozorgnia  \&
  Robertson}{Rahimi et~al.}{2023}]{Rahimi:2022ymu}
Rahimi E.,  Vienneau E.,  Bozorgnia N.,   Robertson A.,  2023, \mn@doi [JCAP]
  {10.1088/1475-7516/2023/02/040}, 02, 040

\bibitem[\protect\citeauthoryear{Ren, Kwa, Kaplinghat  \& Yu}{Ren
  et~al.}{2019}]{Ren:2018jpt}
Ren T.,  Kwa A.,  Kaplinghat M.,   Yu H.-B.,  2019, \mn@doi [Phys. Rev. X]
  {10.1103/PhysRevX.9.031020}, 9, 031020

\bibitem[\protect\citeauthoryear{Robertson, Massey  \& Eke}{Robertson
  et~al.}{2017a}]{Robertson:2016xjh}
Robertson A.,  Massey R.,   Eke V.,  2017a, \mn@doi [Mon. Not. Roy. Astron.
  Soc.] {10.1093/mnras/stw2670}, 465, 569

\bibitem[\protect\citeauthoryear{Robertson, Massey  \& Eke}{Robertson
  et~al.}{2017b}]{Robertson:2016qef}
Robertson A.,  Massey R.,   Eke V.,  2017b, \mn@doi [Mon. Not. Roy. Astron.
  Soc.] {10.1093/mnras/stx463}, 467, 4719

\bibitem[\protect\citeauthoryear{Robertson et~al.}{Robertson
  et~al.}{2018}]{Robertson:2017mgj}
Robertson A.,  et~al., 2018, \mn@doi [Mon. Not. Roy. Astron. Soc.]
  {10.1093/mnrasl/sly024}, 476, L20

\bibitem[\protect\citeauthoryear{Robles, Kelley, Bullock  \& Kaplinghat}{Robles
  et~al.}{2019}]{Robles:2019mfq}
Robles V.~H.,  Kelley T.,  Bullock J.~S.,   Kaplinghat M.,  2019, \mn@doi [Mon.
  Not. Roy. Astron. Soc.] {10.1093/mnras/stz2345}, 490, 2117

\bibitem[\protect\citeauthoryear{Rocha, Peter, Bullock, Kaplinghat,
  Garrison-Kimmel, Onorbe  \& Moustakas}{Rocha et~al.}{2013}]{Rocha:2012jg}
Rocha M.,  Peter A. H.~G.,  Bullock J.~S.,  Kaplinghat M.,  Garrison-Kimmel S.,
   Onorbe J.,   Moustakas L.~A.,  2013, \mn@doi [Mon. Not. Roy. Astron. Soc.]
  {10.1093/mnras/sts514}, 430, 81

\bibitem[\protect\citeauthoryear{Sagunski, Gad-Nasr, Colquhoun, Robertson  \&
  Tulin}{Sagunski et~al.}{2021}]{Sagunski:2020spe}
Sagunski L.,  Gad-Nasr S.,  Colquhoun B.,  Robertson A.,   Tulin S.,  2021,
  \mn@doi [JCAP] {10.1088/1475-7516/2021/01/024}, 01, 024

\bibitem[\protect\citeauthoryear{Sameie, Creasey, Yu, Sales, Vogelsberger  \&
  Zavala}{Sameie et~al.}{2018}]{Sameie:2018chj}
Sameie O.,  Creasey P.,  Yu H.-B.,  Sales L.~V.,  Vogelsberger M.,   Zavala J.,
   2018, \mn@doi [Mon. Not. Roy. Astron. Soc.] {10.1093/mnras/sty1516}, 479,
  359

\bibitem[\protect\citeauthoryear{Sameie, Yu, Sales, Vogelsberger  \&
  Zavala}{Sameie et~al.}{2020}]{Sameie:2019zfo}
Sameie O.,  Yu H.-B.,  Sales L.~V.,  Vogelsberger M.,   Zavala J.,  2020,
  \mn@doi [Phys. Rev. Lett.] {10.1103/PhysRevLett.124.141102}, 124, 141102

\bibitem[\protect\citeauthoryear{Sameie et~al.,}{Sameie
  et~al.}{2021}]{Sameie:2021ang}
Sameie O.,  et~al., 2021, \mn@doi [Mon. Not. Roy. Astron. Soc.]
  {10.1093/mnras/stab2173}, 507, 720

\bibitem[\protect\citeauthoryear{Santos-Santos et~al.,}{Santos-Santos
  et~al.}{2020}]{Santos-Santos:2019vrw}
Santos-Santos I. M.~E.,  et~al., 2020, \mn@doi [Mon. Not. Roy. Astron. Soc.]
  {10.1093/mnras/staa1072}, 495, 58

\bibitem[\protect\citeauthoryear{Spergel \& Steinhardt}{Spergel \&
  Steinhardt}{2000}]{Spergel:1999mh}
Spergel D.~N.,  Steinhardt P.~J.,  2000, \mn@doi [Phys. Rev. Lett.]
  {10.1103/PhysRevLett.84.3760}, 84, 3760

\bibitem[\protect\citeauthoryear{Springel}{Springel}{2005}]{Springel:2005mi}
Springel V.,  2005, \mn@doi [Mon. Not. Roy. Astron. Soc.]
  {10.1111/j.1365-2966.2005.09655.x}, 364, 1105

\bibitem[\protect\citeauthoryear{{Tulin} \& {Yu}}{{Tulin} \&
  {Yu}}{2018}]{2018PhR...730....1T}
{Tulin} S.,  {Yu} H.-B.,  2018, \mn@doi [\physrep]
  {10.1016/j.physrep.2017.11.004}, \href
  {https://ui.adsabs.harvard.edu/abs/2018PhR...730....1T} {730, 1}

\bibitem[\protect\citeauthoryear{Tulin, Yu  \& Zurek}{Tulin
  et~al.}{2013}]{Tulin:2013teo}
Tulin S.,  Yu H.-B.,   Zurek K.~M.,  2013, \mn@doi [Phys. Rev. D]
  {10.1103/PhysRevD.87.115007}, 87, 115007

\bibitem[\protect\citeauthoryear{Turner, Lovell, Zavala  \&
  Vogelsberger}{Turner et~al.}{2021}]{Turner:2020vlf}
Turner H.~C.,  Lovell M.~R.,  Zavala J.,   Vogelsberger M.,  2021, \mn@doi
  [Mon. Not. Roy. Astron. Soc.] {10.1093/mnras/stab1725}, 505, 5327

\bibitem[\protect\citeauthoryear{Vargya, Sanderson, Sameie, Boylan-Kolchin,
  Hopkins, Wetzel  \& Graus}{Vargya et~al.}{2022}]{Vargya:2021qza}
Vargya D.,  Sanderson R.,  Sameie O.,  Boylan-Kolchin M.,  Hopkins P.~F.,
  Wetzel A.,   Graus A.,  2022, \mn@doi [Mon. Not. Roy. Astron. Soc.]
  {10.1093/mnras/stac2069}, 516, 2389

\bibitem[\protect\citeauthoryear{Vogelsberger, Zavala  \& Loeb}{Vogelsberger
  et~al.}{2012}]{Vogelsberger:2012ku}
Vogelsberger M.,  Zavala J.,   Loeb A.,  2012, \mn@doi [Mon. Not. Roy. Astron.
  Soc.] {10.1111/j.1365-2966.2012.21182.x}, 423, 3740

\bibitem[\protect\citeauthoryear{Vogelsberger, Zavala, Simpson  \&
  Jenkins}{Vogelsberger et~al.}{2014}]{Vogelsberger:2014pda}
Vogelsberger M.,  Zavala J.,  Simpson C.,   Jenkins A.,  2014, \mn@doi [Mon.
  Not. Roy. Astron. Soc.] {10.1093/mnras/stu1713}, 444, 3684

\bibitem[\protect\citeauthoryear{Vogelsberger, Zavala, Cyr-Racine, Pfrommer,
  Bringmann  \& Sigurdson}{Vogelsberger et~al.}{2016}]{Vogelsberger:2015gpr}
Vogelsberger M.,  Zavala J.,  Cyr-Racine F.-Y.,  Pfrommer C.,  Bringmann T.,
  Sigurdson K.,  2016, \mn@doi [Mon. Not. Roy. Astron. Soc.]
  {10.1093/mnras/stw1076}, 460, 1399

\bibitem[\protect\citeauthoryear{Wolfram}{Wolfram}{2016}]{wolfram}
Wolfram 2016,
  \url{https://reference.wolfram.com/language/ref/method/GaussianProcess.html}

\bibitem[\protect\citeauthoryear{Xiao, Shen, Hopkins  \& Zurek}{Xiao
  et~al.}{2021}]{Xiao:2021ftk}
Xiao H.,  Shen X.,  Hopkins P.~F.,   Zurek K.~M.,  2021, \mn@doi [JCAP]
  {10.1088/1475-7516/2021/07/039}, 07, 039

\bibitem[\protect\citeauthoryear{Yang \& Yu}{Yang \& Yu}{2021}]{Yang:2021kdf}
Yang D.,  Yu H.-B.,  2021, \mn@doi [Phys. Rev. D]
  {10.1103/PhysRevD.104.103031}, 104, 103031

\bibitem[\protect\citeauthoryear{{Yang} \& {Yu}}{{Yang} \&
  {Yu}}{2022}]{Yang:2022hkm}
{Yang} D.,  {Yu} H.-B.,  2022, \mn@doi [\jcap] {10.1088/1475-7516/2022/09/077},
  \href {https://ui.adsabs.harvard.edu/abs/2022JCAP...09..077Y} {2022, 077}

\bibitem[\protect\citeauthoryear{Yang, Yu  \& An}{Yang
  et~al.}{2020}]{Yang:2020iya}
Yang D.,  Yu H.-B.,   An H.,  2020, \mn@doi [Phys. Rev. Lett.]
  {10.1103/PhysRevLett.125.111105}, 125, 111105

\bibitem[\protect\citeauthoryear{{Yang}, {Jiang}, {Benson}, {Zhong}, {Mace},
  {Du}, {Carton Zeng}  \& {Peter}}{{Yang} et~al.}{2023a}]{Yang:2023stn}
{Yang} S.,  {Jiang} F.,  {Benson} A.,  {Zhong} Y.-M.,  {Mace} C.,  {Du} X.,
  {Carton Zeng} Z.,   {Peter} A. H.~G.,  2023a, \mn@doi [arXiv e-prints]
  {10.48550/arXiv.2305.05067}, \href
  {https://ui.adsabs.harvard.edu/abs/2023arXiv230505067Y} {p. arXiv:2305.05067}

\bibitem[\protect\citeauthoryear{{Yang}, {Nadler}, {Yu}  \& {Zhong}}{{Yang}
  et~al.}{2023b}]{Yang:2023jwn}
{Yang} D.,  {Nadler} E.~O.,  {Yu} H.-B.,   {Zhong} Y.-M.,  2023b, \mn@doi
  [arXiv e-prints] {10.48550/arXiv.2305.16176}, \href
  {https://ui.adsabs.harvard.edu/abs/2023arXiv230516176Y} {p. arXiv:2305.16176}

\bibitem[\protect\citeauthoryear{Yang, Du, Zeng, Benson, Jiang, Nadler  \&
  Peter}{Yang et~al.}{2023c}]{Yang:2022zkd}
Yang S.,  Du X.,  Zeng Z.~C.,  Benson A.,  Jiang F.,  Nadler E.~O.,   Peter A.
  H.~G.,  2023c, \mn@doi [Astrophys. J.] {10.3847/1538-4357/acbd49}, 946, 47

\bibitem[\protect\citeauthoryear{Yang, Nadler  \& Yu}{Yang
  et~al.}{2023d}]{Yang:2022mxl}
Yang D.,  Nadler E.~O.,   Yu H.-B.,  2023d, \mn@doi [Astrophys. J.]
  {10.3847/1538-4357/acc73e}, 949, 67

\bibitem[\protect\citeauthoryear{Zavala, Lovell, Vogelsberger  \&
  Burger}{Zavala et~al.}{2019}]{Zavala:2019sjk}
Zavala J.,  Lovell M.~R.,  Vogelsberger M.,   Burger J.~D.,  2019, \mn@doi
  [Phys. Rev. D] {10.1103/PhysRevD.100.063007}, 100, 063007

\bibitem[\protect\citeauthoryear{Zeng, Peter, Du, Benson, Kim, Jiang,
  Cyr-Racine  \& Vogelsberger}{Zeng et~al.}{2022}]{Zeng:2021ldo}
Zeng Z.~C.,  Peter A. H.~G.,  Du X.,  Benson A.,  Kim S.,  Jiang F.,
  Cyr-Racine F.-Y.,   Vogelsberger M.,  2022, \mn@doi [Mon. Not. Roy. Astron.
  Soc.] {10.1093/mnras/stac1094}, 513, 4845

\bibitem[\protect\citeauthoryear{Zentner, Dandavate, Slone  \& Lisanti}{Zentner
  et~al.}{2022}]{Zentner:2022xux}
Zentner A.,  Dandavate S.,  Slone O.,   Lisanti M.,  2022, \mn@doi [JCAP]
  {10.1088/1475-7516/2022/07/031}, 07, 031

\makeatother
\end{thebibliography}



\appendix

\section{Convergence Tests of N-body simulations}
\label{sec:convergence}

Convergence tests of N-body SIDM simulations in the deep collapse phase are highly nontrivial.~\cite{Yang:2022hkm} found that the number of simulation particles and the time step play essential roles, with details depending on the implementation of dark matter self-interactions. Overall, a smaller time step is favored.~\cite{Yang:2022mxl} further demonstrated that numerical convergence could be achieved for haloes containing fewer than $10^4$ simulation particles if the parameter $\eta$ controlling the time step in~\texttt{GADGET-2} is sufficiently small, i.e.,
\begin{equation}
\Delta t = \sqrt{ \frac{2\eta\epsilon}{|\mathbf{a}|} },
\label{eq:timestep}
\end{equation}
where $\epsilon$ represents the gravitational softening length, and $|\mathbf{a}|$ is the magnitude of a particle's acceleration. In this work, we use the SIDM module developed in~\cite{Yang:2022hkm}. The mass of simulation particles is $3\times10^4~{\rm M_\odot}$, and the total number is $4\times10^6$. For the N-body simulations shown in the main text, we take $\eta=2.5\times 10^{-2}$. We have performed convergence tests for three benchmarks by taking $\eta=2.5\times 10^{-3}$. 

\begin{figure}
	\includegraphics[width=\columnwidth]{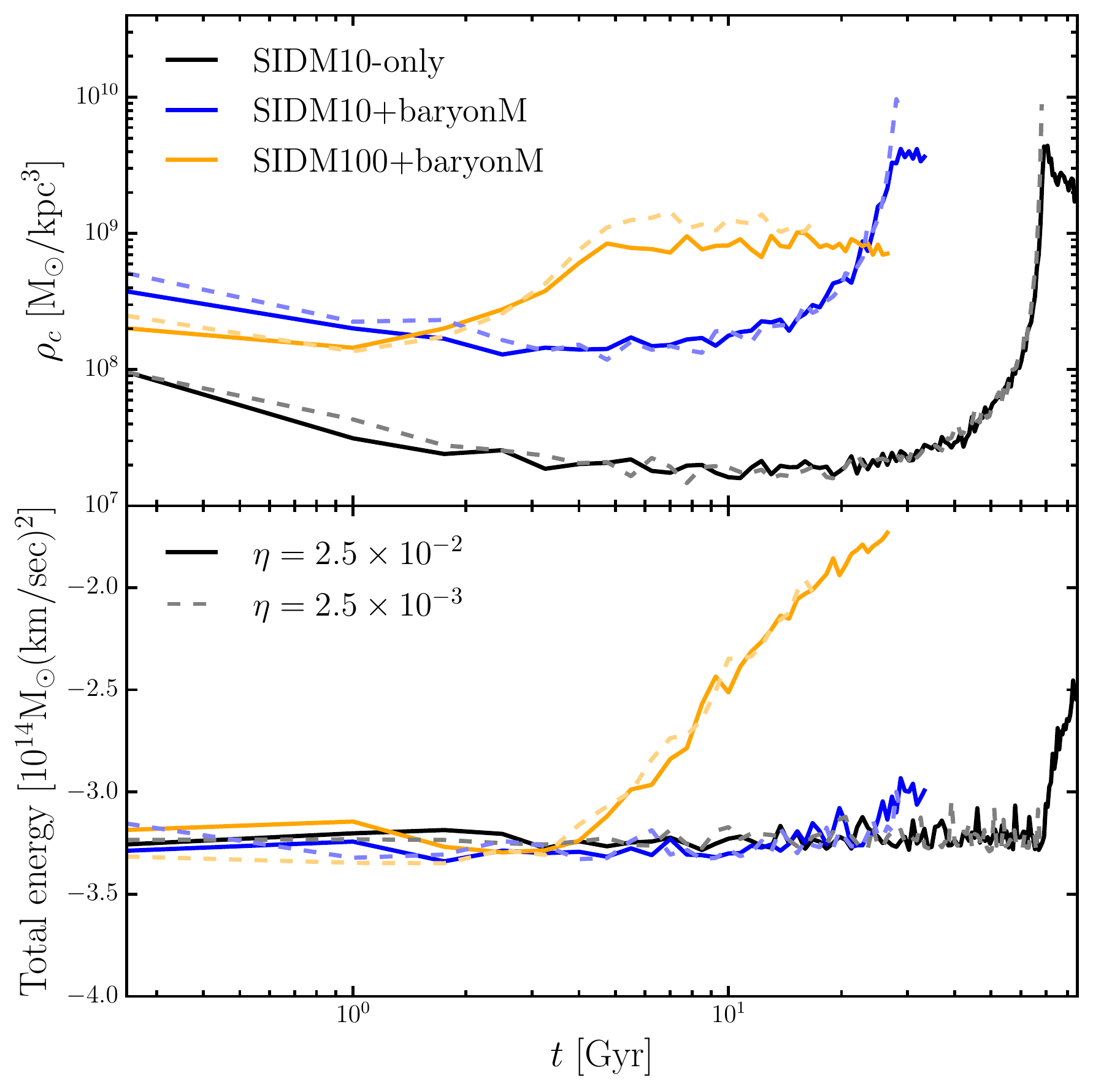}
    \caption{{\bf Top:} Evolution of the central dark matter density for the SIDM10-only (black), SIDM10+baryonM (blue), and SIDM100-baryonM (orange) benchmarks from the N-body simulations, assuming $\eta=2.5\times 10^{-2}$ (solid) and $\eta=2.5\times 10^{-3}$ (dashed).  {\bf Bottom:} Corresponding evolution of the total dark matter energy. The conservation of the total energy serves as a diagnostic indicator for testing the numerical convergence of N-body SIDM simulations in the deep collapse phase. In the main text, we only present the simulation results that the energy conservation holds.}
    \label{fig:conv} 
\end{figure}

Fig.~\ref{fig:conv} (top) shows the evolution of $\rho_c$ the SIDM10-only (black), SIDM10+baryonM (blue), and SIDM100-baryonM (orange) benchmarks from the N-body simulations, assuming $\eta=2.5\times 10^{-2}$ (solid) and $\eta=2.5\times 10^{-3}$ (dashed). The agreement is good before the halo evolves into the deeply collapsed phase, at which the simulated central densities do not increase further. The halt in the growth of the central densities is due to numerical artifacts. This can be seen in Fig.~\ref{fig:conv} (bottom), where we show the corresponding evolution of total dark matter energy of the halo. The total energy increases from its initial value, at which the central density ceases to increase. Since the total energy must be conserved, the ``additional heat'' is artificial and it could be related to the resolution limit, as well as algorithms for modeling dark matter self- and gravitational interactions; see, e.g.,~\cite{Robertson:2016xjh} for related discussion. When we reduce $\eta$ from $2.5\times 10^{-2}$ to $2.5\times 10^{-3}$, the condition improves mildly. In the main text, we present the simulation results that the condition of energy conservation holds and leave further investigation of this topic for future work.

\section{Numerical Recipe of Fluid simulations}
\label{sec:fluidcode}

For the fluid simulation with the baryonic potential, we follow the numerical recipe in~\cite{Feng:2020kxv} to solve~\eqref{fluiddimless}. The halo is segmented to $N=152$ evenly log-spaced concentric Lagrangian zones with radii $\{\hat r_1, \hat r_2, \cdots\ \hat r_N\}$, where $\hat r_1 = 10^{-2}$ and $\hat r_N = 10^3$. The values of extensive quantities $\hat M_i$ and $\hat L_i$ are evaluated at $\hat r_i$ while the intensive quantities $\hat \rho_i$ and $\hat \nu_i$ are taken to be the average between $\hat r_i$ and $\hat r_{i-1}$. We assume that the baryonic potential is static and fix the baryonic mass profile $\hat M_{b}$ as
\beq
\hat M_b (\hat r) = \frac{M_{b,\text{tot}}}{4\pi \rho_{s} r_{s}^3} \left(1 + \frac{r_h}{r_{s}} \hat r^{-1}\right)^{-2}. 
\label{eq:mb}
\eeq

After setting the initial profiles, we conduct the evolution by iterating the ``conduction-then-relaxation'' steps. For a short time interval $\Delta \hat t$, we compute the specific kinetic energy change due to the heat conduction for each Lagrangian zone,
\beq
\frac{3 \Delta \hat \nu^2_{\chi,i}}{2} = -\left(\frac{\partial \hat L_\chi}{\partial \hat M_\chi}\right)_i \Delta \hat t,
\eeq
while keeping the SIDM density $\hat \rho_{\chi,i}$ fixed. We then update $\hat\nu_{\chi,i}^2$ by the resulting amount of $\Delta \hat \nu_{\chi,i}^2$. $\Delta \hat t$ must be sufficiently small, i.e., $|\Delta \hat \nu_{\chi,i}^2/\hat\nu_{\chi,i}^2|\lesssim 10^{-3}$, to guarantee that the linear approximations used in the relaxation step are valid. During this step, $\hat \nu_{\chi,i}$ gets updated, while $\hat \rho_{\chi, i}$ remains the same, the Lagrangian zones are no longer virialized after the conduction, i.e.,
\beq
\left(\frac{\partial \hat p}{\partial \hat r}\right)_i +\frac{( \hat M_i + \hat M_{b,i})\hat \rho_i}{\hat r_i^2} \neq 0.
\eeq
where $\hat p_i = \hat \rho_i \hat \nu_i^2$ and $\hat M_{b,i} = \hat M_b(\hat r_i)$. We have suppressed the superscript ``$\chi$'' for simplicity. The relaxation step gets the zones back to the virial state. The procedure is as follows: we perturb $\hat r_i$, $\hat \rho_{i}$, and $\hat p_{i}$ by a small amount of $\Delta \hat r_i$, $\Delta  \hat \rho_{i}$, and $\Delta \hat p_{i}$, respectively, while keeping the mass $\hat  M_{i}$ and the specific entropy $\hat s_{i}\equiv\ln(\hat \nu_{i}^3/ \hat \rho_{i})$ fixed, to re-establish the hydrostatic equilibrium for all the Lagrangian zones. We obtain the following two linear relations from the conservation laws

\beq
\Delta \hat{\rho}_i = -3\hat{\rho}_i \frac{\hat{r}_i^2 \Delta \hat{r}_i-\hat{r}_{i-1}^2\Delta \hat{r}_{i-1}}{\hat{r}_i^3-\hat{r}_{i-1}^3},\quad
\Delta \hat{p}_i = -5\hat{p}_i \frac{\hat{r}_i^2 \Delta \hat{r}_i-\hat{r}_{i-1}^2\Delta \hat{r}_{i-1}}{\hat{r}_i^3-\hat{r}_{i-1}^3},
\eeq
Substituting them to the linearized perturbed hydrostatic equation, we get a tri-diagonal equation for the perturbation $\Delta\hat r_i$: 
\begin{align}
&\Bigg[\frac{3 (\hat{M}_i+\hat{M}_{b,i}) \hat{\rho}_{i} \hat{r}_{i-1}^2 \hat{r}_{i+1}}{\hat{r}_{i}^3-\hat{r}_{i-1}^3}-\frac{3 (\hat{M}_i+\hat{M}_{b,i}) \hat{\rho}_{i} \hat{r}_{i-1}^3}{\hat{r}_{i}^3-\hat{r}_{i-1}^3}\nonumber\\&-\frac{20 \hat{p}_{i} \hat{r}_{i}^2 \hat{r}_{i-1}^2}{\hat{r}_{i}^3-\hat{r}_{i-1}^3} -(\hat{M}_i+\hat{M}_{b,i}) (\hat{\rho}_{i}+ \hat{\rho}_{i+1})\Bigg] \Delta \hat{r}_{i-1}\nonumber\\
&+\left[-\frac{3 (\hat{M}_i+\hat{M}_{b,i}) \hat{\rho}_{i} \hat{r}_{i}^2 \hat{r}_{i+1}}{\hat{r}_{i}^3-\hat{r}_{i-1}^3}-\frac{3 (\hat{M}_i+\hat{M}_{b,i}) \hat{\rho}_{i+1} \hat{r}_{i}^2 \hat{r}_{i-1}}{\hat{r}_{i+1}^3-\hat{r}_{i}^3}\right.\nonumber\\&+\frac{3 (\hat{M}_i+\hat{M}_{b,i}) \hat{\rho}_{i} \hat{r}_{i}^2 \hat{r}_{i-1}}{\hat{r}_{i}^3-\hat{r}_{i-1}^3}+\frac{3 (\hat{M}_i+\hat{M}_{b,i}) \hat{\rho}_{i+1} \hat{r}_{i}^2  \hat{r}_{i+1}}{\hat{r}_{i+1}^3-\hat{r}_{i}^3}\nonumber\\&\left.+\frac{20  \hat{p}_{i+1}\hat{r}_{i}^4}{\hat{r}_{i+1}^3-\hat{r}_{i}^3}+\frac{20 \hat{p}_{i} \hat{r}_{i}^4}{\hat{r}_{i}^3-\hat{r}_{i-1}^3}+8 \hat{r}_{i} (\hat{p}_{i+1}-\hat{p}_{i})\right] \Delta \hat{r}_{i}\nonumber\\
&+\left[\frac{3 (\hat{M}_i+\hat{M}_{b,i}) \hat{\rho}_{i+1} \hat{r}_{i-1} \hat{r}_{i+1}^2}{\hat{r}_{i+1}^3-\hat{r}_{i}^3}-\frac{3 (\hat{M}_i+\hat{M}_{b,i}) \hat{\rho}_{i+1} \hat{r}_{i+1}^3}{\hat{r}_{i+1}^3-\hat{r}_{i}^3}\right.\nonumber\\
&\left.-\frac{20  \hat{p}_{i+1} \hat{r}_{i}^2 \hat{r}_{i+1}^2}{\hat{r}_{i+1}^3-\hat{r}_{i}^3}+(\hat{M}_i+\hat{M}_{b,i}) (\hat{\rho}_{i}+ \hat{\rho}_{i+1})\right] \Delta \hat{r}_{i+1}\nonumber\\
&+ 4 \hat{r}_{i}^2 \left(\hat{p}_{i+1}-\hat{p}_{i}\right)-(\hat{M}_i+\hat{M}_{b,i}) \left(\hat{\rho}_{i}+\hat{\rho}_{i+1}\right) \left(\hat{r}_{i-1}-\hat{r}_{i+1}\right) =0
\end{align}
After solving $\Delta \hat r_{i}$, we update $\hat r_i$, $\hat \rho_{\chi,i}$, $\hat \nu_{\chi,i}$, and $\hat L_{\chi,i}$ and go back to the conduction step. The evolution is terminated when the Knudsen number for the innermost zone drops far below $0.1$, $Kn_c \ll 0.1$. For SIDM-only simulations, we set $\hat M_{b,i}=0$ and follow the same procedure described above.

\section{Analytical solution to the hydrostatic equation}
\label{sec:hydro}

For the Hernquist baryonic mass profile $\hat M_{b}$ in~\eqref{mb} and the NFW halo mass profile $\hat M_\chi$ at $\hat t=0$ 
\beq
\hat M_\chi (\hat r, \hat t=0)= \ln (1+\hat r) -\frac{\hat r}{1+\hat r},
\eeq
we can analytically solve the hydrostatic equation
\beq
\frac{\partial (\hat \rho_\chi \hat \nu_\chi^2)}{\partial \hat r} = -\frac{(\hat M_\chi + \hat M_b)\hat \rho_\chi}{\hat r^2},
\eeq
to get the 1D velocity dispersion and luminosity profiles. We find that $\hat \nu_{\chi}^2 \equiv \nu_\chi^2/\nu_0^2= \nu_\chi^2/(4\pi G \rho_{s}r_{s}^2)$ can be expressed as
\begin{align}
&\hat \nu_{\chi}^2(\hat r) =\frac{\hat r}{2} \Big\{-\frac{2 \xi^2 \zeta (\hat r+1)}{(\xi-1)^3 (\xi \hat r+1)} \Big[-\xi^4 \hat r^2 \ln (\xi \hat r+1)\nonumber\\
&+3 \xi^3 \hat r^2 \ln (\xi \hat r+1)+\xi^3 \hat r+2 \xi^3 \hat r \ln (\xi \hat r+1)-\xi^3 \ln (\xi \hat r+1)+\xi^3\nonumber\\
&-3 \xi^2 \hat r^2 \ln (\hat r+1)+(\xi-3) \xi^2 (\hat r+1) (\xi \hat r+1) \ln (\xi)-3 \xi^2 \hat r \ln (\hat r+1)\nonumber\\
&+3 \xi^2 \hat r \ln (\xi \hat r+1)+3 \xi^2 \ln (\xi \hat r+1)-\xi^2+\xi \hat r^2 \ln (\hat r+1)-\xi \hat r\nonumber\\
&-2 \xi \hat r \ln (\hat r+1)-3 \xi \ln (\hat r+1)+(\xi-1)^3 (\hat r+1) (\xi \hat r+1) \ln (\hat r)\nonumber\\
&+\xi+\hat r \ln (\hat r+1)+\ln (\hat r+1)-1-\xi^4 \hat r \ln (\xi \hat r+1)\Big] \nonumber\\
&+\frac{(\hat r+1)}{\hat r^2} \Big[6 (\hat r+1) \hat r^2 \text{Li}_2(-\hat r)+\pi ^2 \hat r^3+3 \hat r^3 \ln ^2(\hat r+1)-5 \hat r^3 \ln (\hat r+1)\nonumber\\
&+\pi ^2 \hat r^2-\hat r^2+3 \hat r^2 \ln ^2(\hat r+1)+5 (\hat r+1) \hat r^2 \ln (\hat r)-11 \hat r^2 \ln (\hat r+1)+\hat r\nonumber\\
&-3 \hat r \ln (\hat r+1)+\ln (\hat r+1)\Big]\nonumber\\
&+\frac{1}{\hat r}\left[-6 \hat r^2-9 \hat r+6 (\hat r+1)^2 \hat r \ln \left(\hat r^{-1}+1\right)-2\right]\Big\},
\label{eq:nut0}
\end{align}
where $\xi \equiv r_{s}/r_{h}$ and $\zeta\equiv M_{b,\text{tot}}/(4\pi \rho_{s} r_{s}^3)$ and Li$_2$ is the polylog function. When $\zeta= 0$, we obtain the 1D velocity dispersion for an NFW profile.

\section{Collapse time}
\label{sec:linear_relation}

\begin{figure}
    \centering
    \includegraphics[width=0.89\columnwidth]{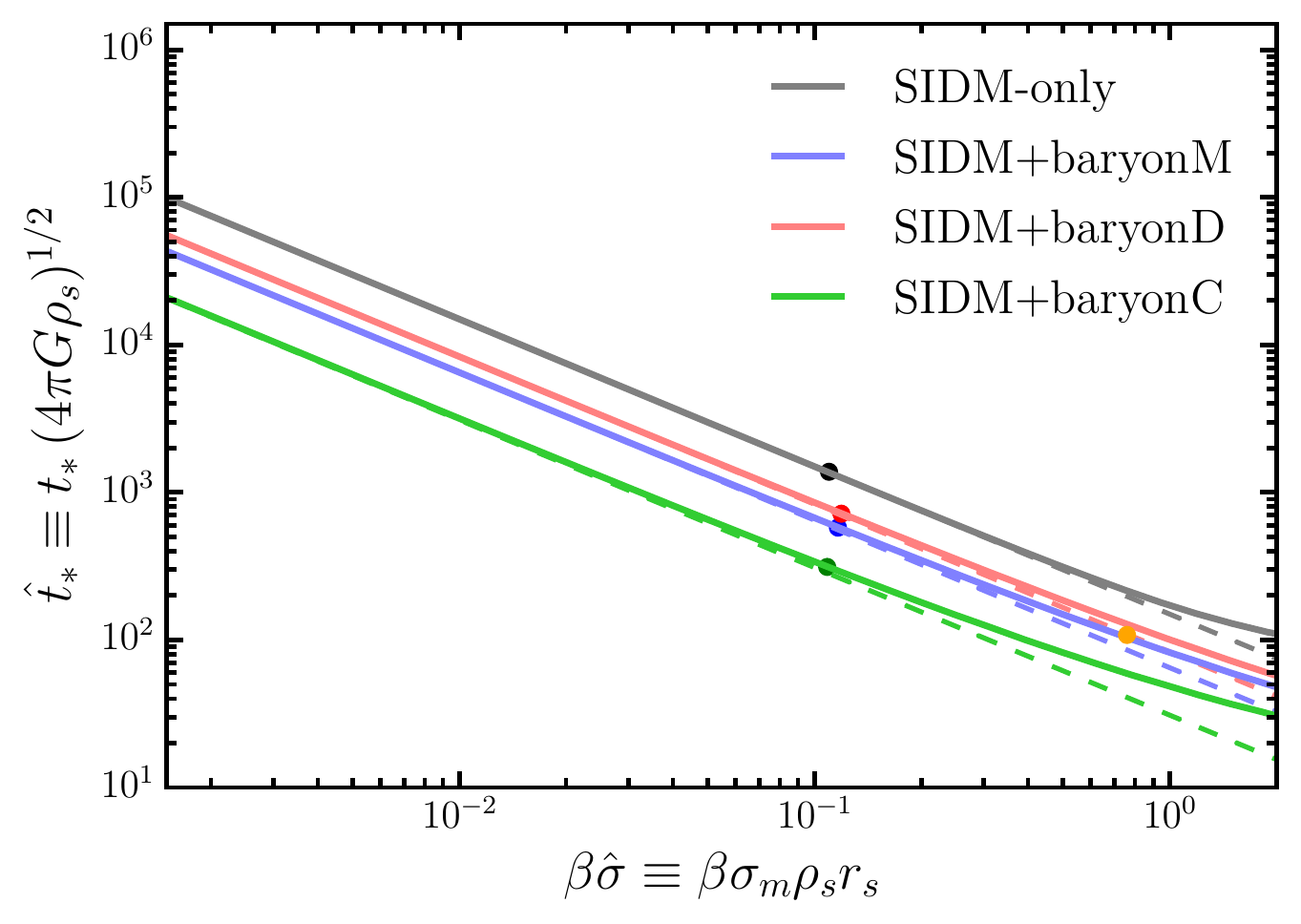}
    \caption{The dimensionless collapse time $\hat t$ as a function of $\beta \hat \sigma$ from the fluid simulations for SIDM-only, as well as three SIDM+baryon configurations (solid). Their corresponding predictions from~\eqref{newfit} are also shown (dashed) for comparison. The circles represent the benchmarks listed in Table~\ref{tab:benchmark}, i.e., SIDM10-only (black), SIDM10+baryonM (blue), +baryonD (red), and +baryonC (green), as well as SIDM100+baryonM (yellow).}
    \label{fig:collapse_time}
\end{figure}

\figref{collapse_time} shows the dimensionless collapse time $\hat t$ for the SIDM-only case, as well three SIDM+baryon configurations, as a function of $\beta \hat \sigma_m$ from the fluid simulations (solid). For comparison, we also plot 
\beq
t_*  = \frac{\gamma}{\beta\sigma_m^\text{eff}} \frac{1}{ \rho_{s} r_{s}} \frac{1}{\sqrt{4\pi G \rho_{s}}}
\label{eq:newfit}
\eeq
for each case (dashed), where $\gamma = \{150, 65, 83.3, 31\}$ for SIDM-only, SIDM+baryonM, D, and C, respectively. We see that the scaling relation $\hat t_* \propto (\beta\hat\sigma)^{-1}$ largely holds. However, if $\sigma_m$ is large enough, the deviation occurs, and the collapse time is longer than predicted in \eqref{newfit}. In this case, the conductivity $\kappa$ is no longer solely determined by $\kappa_\text{lmfp}$, and one needs to include $\kappa_\text{smfp}$ contributions, which do not scale with $\sigma_m$. The critical $\beta\sigma_m$ value at which the relation deviates is correlated with the compactness of the baryonic potential. As the potential deepens, the value decreases because the velocity dispersion increases accordingly, and the short-mean-free-path condition can be satisfied easier ($K_n < 1$) as $Kn \propto \nu_\chi^{-1}$. In~\figref{collapse_time}, we also show the five constant $\sigma_m$ benchmarks listed in Table~\ref{tab:benchmark} (colored circle). For SIDM100+baryonM (yellow circle), the relation $\hat t_* \propto (\beta\hat\sigma)^{-1}$ is violated mildly.

\section{Gaussian Process Regression}
\label{sec:nbodyextract}

Gaussian Process Regression (GPR) has been widely used for analyzing astronomical time-series data~\citep{2022arXiv220908940A}. Since the method is simple, flexible, and robust, it is an ideal tool for modeling stochastic signals in such data. In our work, we utilize GPR to analyze the temporal evolution of $\rho_c$ or $r_c$ from the N-body simulations, where numerical fluctuations are present; see~\figref{densitytime1}. We first use GPR to fit the simulation data for $\rho_c(t)$ and $r_c(t)$. Then, we determine $t_m$, as well as $\rho_c(t_m)$ and $r_c(t_m)$, based on the fits.

\begin{figure*}
    \centering
    \includegraphics[width=\textwidth]{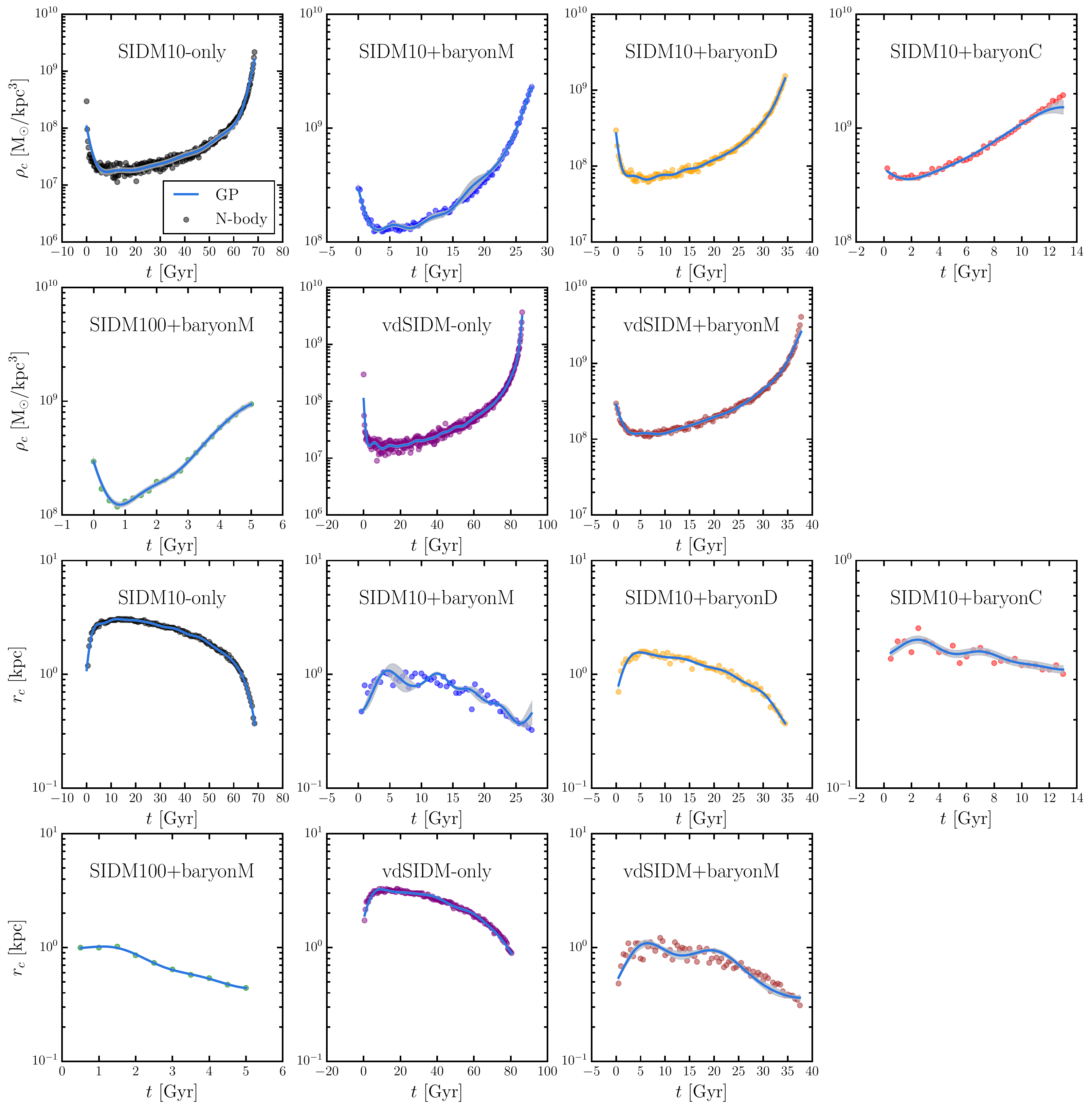}
    \caption{Evolution $\rho_c(t)$ and $r_c(t)$ from the GPR fits with the mean value (solid blue) and the $\pm 1 \sigma$ range (gray band), compared to the N-body simulations (colored circle).
   }
    \label{fig:GP}
\end{figure*}

We use the~\texttt{Predict} module of \texttt{Mathematica} 13 with the~\texttt{GaussianProcess} method~\citep{wolfram}, and choose the squared exponential kernel as the covariance function. We perform GPR fits for the $\rho_c(t)$ and $r_c(t)$ data from the N-body simulations of the benchmarks listed in Table~\ref{tab:benchmark}. \figref{GP} shows the mean value (solid blue), and the $\pm 1 \sigma$ range (gray band) from the GPR fits, compared to the N-body simulations (colored dots). From the fits, we can uniquely determine the moment when $\rho_c$ ($r_c$) reaches minimal (maximal) for each benchmark. If the $t_m$ value extracted from $\rho_c$ does not coincide with that from $r_c$, we report both values; see Table~\ref{tab:extract}.


\bsp	
\label{lastpage}
\end{document}